\documentclass{ieeeaccess}
\usepackage{cite}
\usepackage{amsmath,amssymb,amsfonts}
\usepackage{algorithmic}
\usepackage{graphicx}
\usepackage{textcomp}
\def\BibTeX{{\rm B\kern-.05em{\sc i\kern-.025em b}\kern-.08em
    T\kern-.1667em\lower.7ex\hbox{E}\kern-.125emX}}
    
\usepackage{float}
\usepackage[hyphens]{url}
\usepackage[breaklinks=true,colorlinks=true,bookmarks=false]{hyperref}
\hypersetup{pdfauthor={Mohaimen}}
\usepackage{pgfplots}
\pgfplotsset{compat=1.8}
\usepackage[caption=false]{subfig}

\pgfplotsset{
    bar group size/.style 2 args={
        /pgf/bar shift={%
                -0.5*(#2*\pgfplotbarwidth + (#2-1)*\pgfkeysvalueof{/pgfplots/bar group skip})  + 
                (.5+#1)*\pgfplotbarwidth + #1*\pgfkeysvalueof{/pgfplots/bar group skip}},%
    },
    bar group skip/.initial=6pt,
    plot 0/.style={black,fill=black!70,mark=none},%
    plot 1/.style={black,fill=purple,mark=none},%
    plot 2/.style={black,fill=orange,mark=none},%
}

\newcommand{\ignore}[2]{\hspace{0in}#2}
\NewSpotColorSpace{PANTONE}
\AddSpotColor{PANTONE} {PANTONE3015C} {PANTONE\SpotSpace 3015\SpotSpace C} {1 0.3 0 0.2}
\SetPageColorSpace{PANTONE}%
\definecolor{accessblue}{cmyk}{1 0.3 0 0.2}%
\definecolor{greycolor}{cmyk}{0,0,0,.8}

\usetikzlibrary{shapes,arrows.meta,fit}
\definecolor{myblue1}{RGB}{218, 232, 252}
\definecolor{myblue2}{RGB}{108, 142, 191}

\begin{document}
\history{Received November 25, 2021, accepted December 27, 2021}
\doi{10.1109/ACCESS.2022.3140807}

%\title{Pruning vs XNOR-Net: Which One to Go for Audio Classification in Micro-Controllers?}
\title{Pruning vs XNOR-Net: A Comprehensive Study of Deep Learning for Audio Classification on Edge-devices}
\author{\uppercase{Md Mohaimenuzzaman}\authorrefmark{1}, \uppercase{Christoph Bergmeir\authorrefmark{1} and Bernd Meyer}\authorrefmark{1}}
\address[1]{Department of Data Science and AI, Monash University, Australia (e-mail: md.mohaimen, christoph.bergmeir, bernd.meyer@monash.edu)}

\tfootnote{This work was supported in part by the Australian Research Council under grant DE190100045, in part by Monash University Graduate Research Funding, and in part by the MASSIVE - high performance computing facility, Australia.}

\markboth
{mohaimenuzzaman \headeretal: Pruning vs XNOR-Net: A Comprehensive Study of Deep Learning}
{mohaimenuzzaman \headeretal: Pruning vs XNOR-Net: A Comprehensive Study of Deep Learning}

\corresp{Corresponding author: Md Mohaimenuzzaman(e-mail: md.mohaimen@monash.edu).}

\begin{abstract}
Deep learning has celebrated resounding successes in many application areas of relevance to the Internet of Things (IoT), such as computer vision and machine listening. These technologies must ultimately be brought directly to the edge to fully harness the power of deep leaning for the IoT. The obvious challenge is that deep learning techniques can only be implemented on strictly resource-constrained edge devices if the models are radically downsized. This task relies on different model compression techniques, such as network pruning, quantization, and the recent advancement of XNOR-Net. This study examines the suitability of these techniques for audio classification on microcontrollers. We present an application of XNOR-Net for end-to-end raw audio classification and a comprehensive empirical study comparing this approach with pruning-and-quantization methods. We show that raw audio classification with XNOR yields comparable performance to regular full precision networks for small numbers of classes while reducing memory requirements 32-fold and computation requirements 58-fold. However, as the number of classes increases significantly, performance degrades, and pruning-and-quantization based compression techniques take over as the preferred technique being able to satisfy the same space constraints but requiring approximately 8x more computation. We show that these insights are consistent between raw audio classification and image classification using standard benchmark sets. To the best of our knowledge, this is the first study to apply XNOR to end-to-end audio classification and evaluate it in the context of alternative techniques. All codes are publicly available on GitHub.

\end{abstract}


\titlepgskip=-15pt

\maketitle

\section{Introduction}
\label{sec:introduction}
%Make all the math bold. To make normal for a particular place use \unboldmath
\boldmath
\PARstart{A}{udio} classification is a fundamental building block of many smart Internet of Things (IoT) applications such as predictive maintenance~\cite{der2019PredMaintenance, jia2018PredMaintenance, yun2020PredMaintenance}, surveillance~\cite{sharan2016OverviewSoundRecognition, xu2020CNNAecAnimalSpeech}, and ecosystem monitoring~\cite{stowell2019SoundWildEcoMonitoring, yan2019SoundWildEcoMonitoring}. Smart sensors driven by microcontroller units (MCUs) are at the core of these applications. MCUs, installed at the edge of the networks, sense data and send them to the cloud for classification and detection. However, the energy requirements for transmitting high volumes of data are a burden for battery-powered MCUs. Furthermore, this increases latency, and data transmission may further lead to privacy concerns. The increased latency makes real-time or near real-time analytics infeasible. One way to solve these challenges is to move the analysis and recognition directly to the edge. In practice, this means that all the processing must take place on resource-impoverished MCUs.

MCUs are low-powered resource-constrained devices typically based on system-on-a-chip (SoC) hardware with less than a megabyte (1 MB) of RAM and below 200 MHz clock speeds. All the recent state-of-the-art audio classification models, however, are based on deep learning (DL)~\cite{gong2021AudioSpectroTransformer, nanni2021EnsenbleCNNAudio, mohaimen2021ACDNet, kim2020MultiChannel, kumar2020SeqSelfTeaching, zhang2019OnESC, su2019Urbansound8k, tokozume2017Envnet2, sailor2017UnSupervisedFilterbank} requiring very resource-intensive computation. Usually, the memory size of such models varies from several MBs to even gigabytes (GB). To run such models on MCUs requires extreme minimization of the model's size and computation requirements with minimum loss of accuracy. Recent studies have applied model compression techniques such as pruning connections and neurons from fully connected neural networks (FCNNs)~\cite{han2015DeepCompression}, filter or channel pruning from convolutional neural networks (CNNs)~\cite{structuredpruning, molchanov2016PrunningCNNNvdia, oyedotun2020SCompLasso}, knowledge distillation~\cite{hinton2015distilling} and low-precision quantization \cite{han2015DeepCompression, polino2018CompViaDistQuan}. The most recent advancement in extreme downsizing of DL models and their computation requirements is XNOR-Net~\cite{rastegari2016XNOR}, where a model's activations and inputs are fully binarized.

There are many state-of-the-art XNOR-Net models for different computer vision tasks, such as Rastegari et al.~\cite{rastegari2016XNOR} for MNIST, Cong~\cite{congXNORNet} for CIFAR-10, and Bulat et al.~\cite{bulat2020ExpertBinNet} for CIFAR-100 and ImageNet datasets. However, for audio classification, the only work we are aware of is presented by Cerutti et al.~\cite{cerutti2020AudioXnor}. This study uses XNOR-Net in combination with spectrogram-based input to effectively perform audio classification via image classification. As the literature on full-precision audio networks shows, this approach does not usually deliver the best results for difficult classification problems with conventional CNN architectures\cite{rothmann2018WhatsWrongWithSpectroAsImage, wyse2017SpectroByCNN} (see Section~\ref{sec:handling-audio-data} for further details). To the best of our knowledge, the current literature has not yet considered XNOR-Net for raw audio classification.

Most of the work on XNOR has been performed for computer vision. The leaderboards of benchmark image datasets show considerable differences in the classification accuracy of state-of-the-art full precision nets and XNOR-Nets (see Table~\ref{tab:sota-image} for the CIFAR-100 and ImageNet datasets). Furthermore, models produced by XNOR-Net generally yield a up to $32x$ reduction in memory requirements and up to $58x$ reduction in computation cost compared to their full-precision counterparts~\cite{rastegari2016XNOR}. This may not result in sufficient reduction for MCUs. The memory requirements of XNOR-Net-based DL models producing comparable accuracy \cite{rastegari2016XNOR, congXNORNet, bulat2020ExpertBinNet} typically reach several MBs, while typical MCUs offer only 128KB to 1MB of memory. 

In this paper, we seek to understand the comparison of XNOR model minimization with pruning-and-quantization approaches and the potential of XNOR in the context of audio classification. We present  XNOR-Nets for raw audio classification followed by a comprehensive study that compares this approach with  traditional model compression techniques (pruning-and-quantization). Our extensive experimental study reveals that XNOR-Net may be preferred for scenarios comprising a small number of classes (e.g., 10 classes) along with extremely tight resource constraints, specifically where computation speed is concerned. In contrast, for complex scenarios with a larger number of classes, pruning-and-quantization-based compression techniques would still be the choice because they produce sufficiently small models for off-the-shelf MCUs that have higher performance in terms of accuracy.

Experiments show that, when two models are generated by pruning-and-quantization and XNOR-Net for the same memory constraint, XNOR-Net requires $\sim\!\!8x$ less computation while both produce comparable classification accuracy for small problem sizes (number of classes). While XNOR-Nets require extremely little computation compared to their pruning-and-quantization based counterparts, their classification performance on datasets degrades rapidly with an increasing number of classes. The performance loss of pruning-and-quantization based models for the same audio benchmarks is much more graceful, so that this approach still appears to be preferable for problems with larger class numbers.

To harden the above findings, we have conducted a similar study on image classification datasets \ignore{(namely CIFAR-10 and CIFAR-100) using RESNET-18}  and analyzed this in the context of current state-of-the-art full precision and XNOR-Net leaderboards for various image benchmarks. The behavior is found to be consistent in both the audio and the image domains.

Thus, the contribution of this paper is twofold: 1) it presents the first application of XNOR-Net for raw audio classification as a benchmark for future research. 2) It presents the first comprehensive empirical comparison of pruning, quantization, and XNOR-Net-based model compression techniques and derives guidelines for when to use pruning, quantization, and XNOR-Net, respectively. 

\section{Background and Related Work}
\label{sec:bg-rw}
\subsection{Audio Representations}
\begin{figure}[H]
  \centering   
  \fbox{
  \includegraphics[width=0.75\columnwidth, height=4cm]{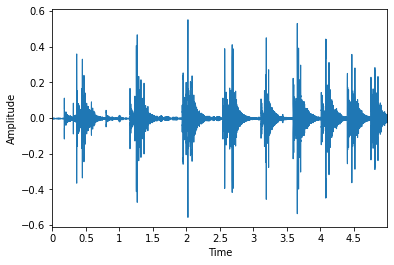}} 
  \caption{1-D representation of audio as wave}
  \label{fig:audio1D}
\end{figure}

Audio can be represented in the time domain as a wave form of amplitude changes over time (Figure~\ref{fig:audio1D}).  If this time series is directly used as an input to the network, we speak of raw audio classification. 

\begin{figure}[H]
   \centering
   \fbox{
   \includegraphics[width=0.75\columnwidth, height=4cm]{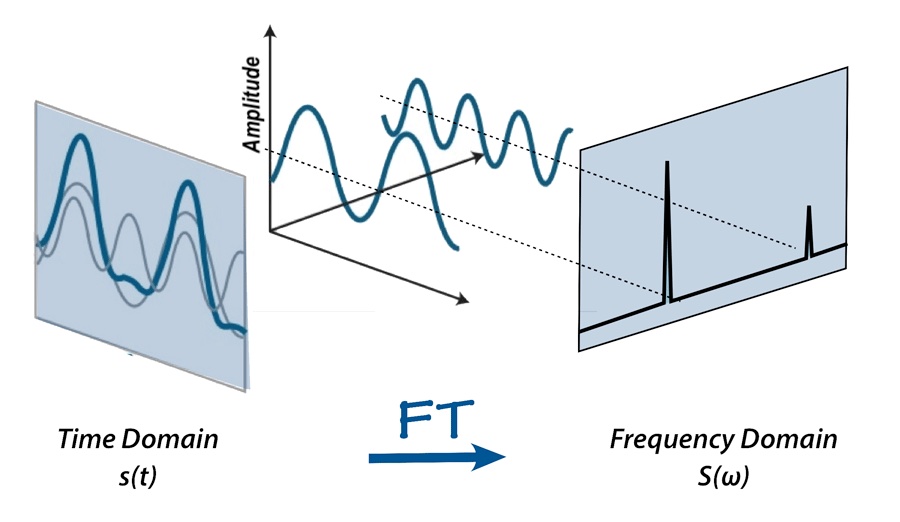}}  
   \caption{2-D representation of audio \cite{chaudhary2021FFT}.}
   \label{fig:audio2D}
\end{figure}

\begin{figure}[H]
   \centering
   \fbox{
   \includegraphics[width=0.75\columnwidth, height=4.5cm]{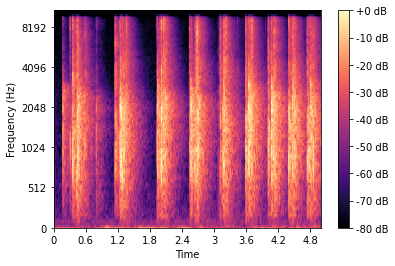}}  
   \caption{Example spectrogram}
   \label{fig:sound-as-image}
\end{figure}

An alternative is to use spectrograms. For this, the audio is first transformed into the frequency domain using Fourier transforms of short overlapping windows and presented with respect to time, frequency, and amplitude (Figure~\ref{fig:audio2D}). The spectrogram captures the intensity of the different frequency component of the signal against time (see Figure~\ref{fig:sound-as-image}).

\subsection{Audio Feature Processing}
\label{sec:handling-audio-data}
Input representation is a fundamental decision when applying deep learning to any problem. Owing to the tremendous success of deep learning in the image processing domain,  researchers have used spectrograms directly as the input representations~\cite{verma2018SpectroAsImage, wyse2017SpectroByCNN, irfan2021deepship, ekpezu2021aecMfcc}. Somewhat surprisingly, the results achieved so far have not matched the performance that might have been expected based on the state-of-the-art in visual image processing~\cite{rothmann2018WhatsWrongWithSpectroAsImage, wyse2017SpectroByCNN}. 

While being a visual structure, spectrograms have properties different from those of natural images. A pixel of a certain color or the similar neighboring pixels of an image may often belong to the same visual object. On the other hand, although frequencies move together according to a common relationship of the sound, a particular frequency or several frequencies do not belong to a single sound \cite{wyse2017SpectroByCNN, rothmann2018WhatsWrongWithSpectroAsImage}. Furthermore, in images, both axes carry spatial information, but the axes of the spectrograms have different meanings. Sounds are not static two-dimensional objects such as images; they genuinely are time series. Thus, the invariances in the natural images and spectrograms are fundamentally different. For example, moving a face in any direction does not change the fact that it is the same face. However, moving frequencies upwards may not only change an adult voice to a child's voice or something completely different, it may also change the spatial information of the sound \cite{wyse2017SpectroByCNN}. 

Hence, this research considers audio classification using raw audio time series, an approach that has proven to be successful with traditional full-precision networks~\cite{mohaimen2021ACDNet}.

\subsection{Audio Classification at the Edge}
The recent state-of-the-art performances for audio classification are all produced by different resource-intensive DL models~\cite{mohaimen2021ACDNet, zhang2019OnESC, su2019Urbansound8k, tokozume2017Envnet2, sailor2017UnSupervisedFilterbank}. These models must be extremely compressed to be run on the MCUs.

\subsubsection{Model Compression}
There are different DL model compression techniques such as unstructured compression ~\cite{han2015DeepCompression} (i.e. weight pruning), structured compression~\cite{molchanov2016PrunningCNNNvdia, oyedotun2020SCompLasso} (i.e. channel/filter/neuron pruning), knowledge distillation~\cite{hinton2015distilling}, and quantization~\cite{polino2018CompViaDistQuan}.

Unstructured pruning produces sparse weight matrices that requires sparse computation to fully utilize its benefits. This is not yet supported by the MCUs~\cite{anwar2017StructuredPruningCNN, gordon2018MorphNet, li2016PruningFiltersConvNet, luo2018ThiNet, singh2020SCompFilterCorel}. 

In structured pruning, the neurons and the channels are pruned to find a tiny version of the baseline network. This is an iterative process in which a single iteration performs a global ranking of filters/neurons (using methods such as L2-Norm, Taylor criteria~\cite{molchanov2016PrunningCNNNvdia} and binary index-based ranking~\cite{luo2020autopruner}), followed by the removal of the lowest-ranked neuron/channel from the network and retraining of the network for one or two epochs to recover the loss of accuracy due to the pruning~\cite{molchanov2016PrunningCNNNvdia}. This is repeated until the target number of filters/neurons is removed from the network to find a model of the desired size. This process is illustrated in Figure~\ref{fig:iterative_structured_pruning}. 

% Define block styles
\tikzstyle{io} = [trapezium, draw=myblue2, fill=myblue1, text width=5em, text centered, trapezium left angle=75, trapezium right angle=105]
\tikzstyle{process} = [rectangle, draw=myblue2, thin, fill=myblue1, text width=10em, text centered]
\tikzstyle{decision} = [diamond, draw=myblue2, fill=myblue1, text width=3em, text badly centered, inner sep=0pt]
\tikzstyle{connect} = [draw, thick, -Latex]%[draw, -Latex]
\tikzstyle{border} = [rectangle, draw, solid, thin, black]
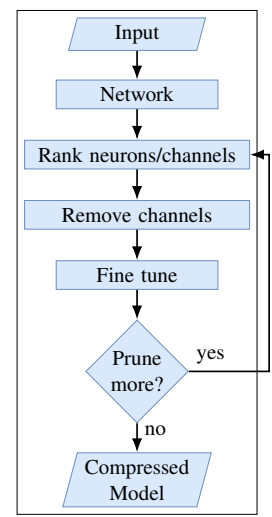
\begin{figure}[H]
	\centering
% 	\begin{center}
	\resizebox{0.4\columnwidth}{!}{%
    \begin{tikzpicture}
    	%Left Nodes
    	\node(start)[io] {Input};
    	\node(step1)[process, below of=start, node distance=10mm, text width=7em] {Network};
    	\node(step2)[process, below of=step1, node distance=10mm] {Rank neurons/channels};
    	\node(step3)[process, below of=step2, node distance=10mm] {Remove channels};
    	\node(step4)[process, below of=step3, node distance=10mm, text width=7em] {Fine tune};
    	\node(step5)[decision, below of=step4, node distance=16mm] {Prune more?};
    	\node(end)[io, below of=step5, node distance=18mm] {Compressed Model};
    	
    	\path[connect] (start.south) -- (step1.north);
    	\path[connect] (step1.south) -- (step2.north);
    	\path[connect] (step2.south) -- (step3.north);
    	\path[connect] (step3.south) -- (step4.north);
    	\path[connect] (step4.south) -- (step5.north);
    	\path[connect] (step5.east) node[above right]{yes} -| (2.2, -2) |- (step2.east);
    	\path[connect] (step5.south) -- node[near start, right]{no} (end.north);
        \node(border) [border, fit=(start)(step1)(step2)(step3)(step4)(step5)(end)]  {};
    \end{tikzpicture}
    }
    % \end{center}
    % \vspace{-3mm}
	\caption{Iterative process of structured model compression}
	\label{fig:iterative_structured_pruning}
\end{figure}

There has been a significant amount of research on structured channel pruning, such as \cite{molchanov2016PrunningCNNNvdia, oyedotun2020SCompLasso, anwar2017StructuredPruningCNN, gordon2018MorphNet, luo2018ThiNet, singh2020SCompFilterCorel, mohaimen2021ACDNet}. However, all of this except \cite{mohaimen2021ACDNet} is proposed for computer vision. The hybrid channel pruning technique proposed in \cite{mohaimen2021ACDNet} is the only study to date for audio analysis. This method incorporates the Taylor expansion criteria (TE)~\cite{molchanov2016PrunningCNNNvdia} in the ranking process. In the beginning of an iteration, a forward pass with the entire training dataset takes place and the gradient is applied to the activations to determine the least affected channels for ranking. The change in gradient can be expressed as: 

\begin{equation}
    \Theta_{TE}(Z_{l}^{i}) = \mid \Delta C Z_{l}^{i} \mid
    \label{eqn:prune-taylor-criteria}
\end{equation}
where $Z_{l}^{i}$ is the $i$th feature map of layer $l$, $\Delta{C}$ is the change in loss denoted by $\frac{\delta C}{\delta Z_{l}^{i}}$ and $\Theta_{TE}(Z_{l}^{i})$ denotes the change determined by TE. Now, the gradient is applied to the activation as:
\begin{equation}
    Z_{l}^{i} = Z_{l}^{i} + \Theta_{TE}(Z_{l}^{i})
    \label{eqn:prune-taylor-activation}
\end{equation}

\noindent
For the ranking of the channels, all the feature maps are normalized layer-wise. Thus, the normalization for a layer $l$ of a network can be expressed as:
\begin{equation}
    \bar{Z_{l}} = \frac{|Z_{l}^{(i)}|}{\sqrt{\sum{|Z_{l}|^2}}}
    \label{eqn:gen-layer-norm}
\end{equation}
where $Z_{l}$ is the list of activations for all the channels $c$ of a layer $l$ in a CNN. Now, the ranking of the channels across all layers is performed and the index of the lowest-ranked channel is determined as follows:
\begin{equation}
    i_{lc} = \kappa(\bar{Z})
    \label{eqn:gen-rank}
\end{equation}
where $\bar{Z}$ are the normalized activations of all  channels of all  layers, $\kappa$ is the function that takes $\bar{Z}$ and returns the information of the channel with the lowest magnitude $i_{lc}$, where $i$ is the index of channel $c$ of layer $l$. 

Once this iterative process produces the final compressed and fine-tuned model, it is further compressed using low-precision quantization. Quantization is an independent process, and this study uses 8-bit quantization to achieve a further $4x$ compression. 

Knowledge distillation is a very different approach to pruning where the knowledge of a large teacher network is gradually transferred to a smaller student network. However, \cite{mohaimen2021ACDNet} showed that structured pruning produces superior performance to knowledge distillation in audio classification tasks. 

According to the current literature, the best pruning-based approach to derive a tiny model from a state-of-the-art deep neural network (DNN) model is structured pruning. Furthermore, the only work that successfully deploys a DNN model for audio classification to MCUs uses this technique followed by 8-bit post-training quantization to compress the state-of-the-art model \cite{mohaimen2021ACDNet}. In that study, the compressed model still produces close to that of the state-of-the-art classification accuracy. Thus, for DNN models using pruning-based techniques, this study focuses on structured pruning and quantization. For simplicity, we refer to this as ``pruning-and-quantization''.

\subsubsection{XNOR-Net}
\label{sec:bg-xnor}
In an XNOR-Net~\cite{rastegari2016XNOR}, all layers except the first and the last are binary. The input, activations, and the weights of the binary layers are represented using either +1 or -1 and are stored efficiently with single bits. Figure~\ref{fig:conv_vs_bin_conv} shows the construction of a typical convolution layer and an equivalent binary convolution layer. 

% Define block styles
\tikzstyle{layer-box} = [rectangle, draw=myblue2, double, thin, fill=myblue1, text width=7em, text centered]
\tikzstyle{border-box} = [rectangle, draw, solid, thin, black]
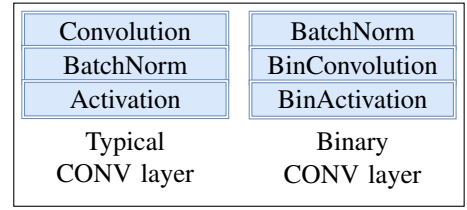
\begin{figure}[H]
	\centering
	\begin{center}
    \begin{tikzpicture}
    	%Left Nodes
    	\node(fullConv)[layer-box] {Convolution};
    	\node(fullBN)[layer-box, below of=fullConv, node distance=4.5mm] {BatchNorm};
    	\node(fullActivation)[layer-box, below of=fullBN, node distance=4.5mm] {Activation};
    	\node (label1)[layer-box, below of=fullActivation, draw=white, node distance=8mm, fill=white]{Typical CONV layer};
    	
    	%Right Nodes
    	\node(binBN)[layer-box, right of=fullConv, node distance=3cm] {BatchNorm};
    	\node(binConv)[layer-box, below of=binBN, node distance=4.55mm] {BinConvolution};
    	\node(binActivation)[layer-box, below of=binConv, node distance=4.5mm] {BinActivation};
    	\node (label2)[layer-box, below of=binActivation, draw=white, node distance=8mm, fill=white]{Binary CONV layer};
    	
    	\node(border) [border-box, fit=(fullConv)(fullBN)(fullActivation)(label1)(binBN)(binConv)(binActivation)(label1)]  {};
    \end{tikzpicture}
    \end{center}
    \vspace*{-3mm}
	\caption{Typical convolution layer vs binary convolution layer}
	\label{fig:conv_vs_bin_conv}
\end{figure}

\begin{figure}[H]
   \centering
   \fbox{
   \includegraphics[width=0.95\columnwidth]{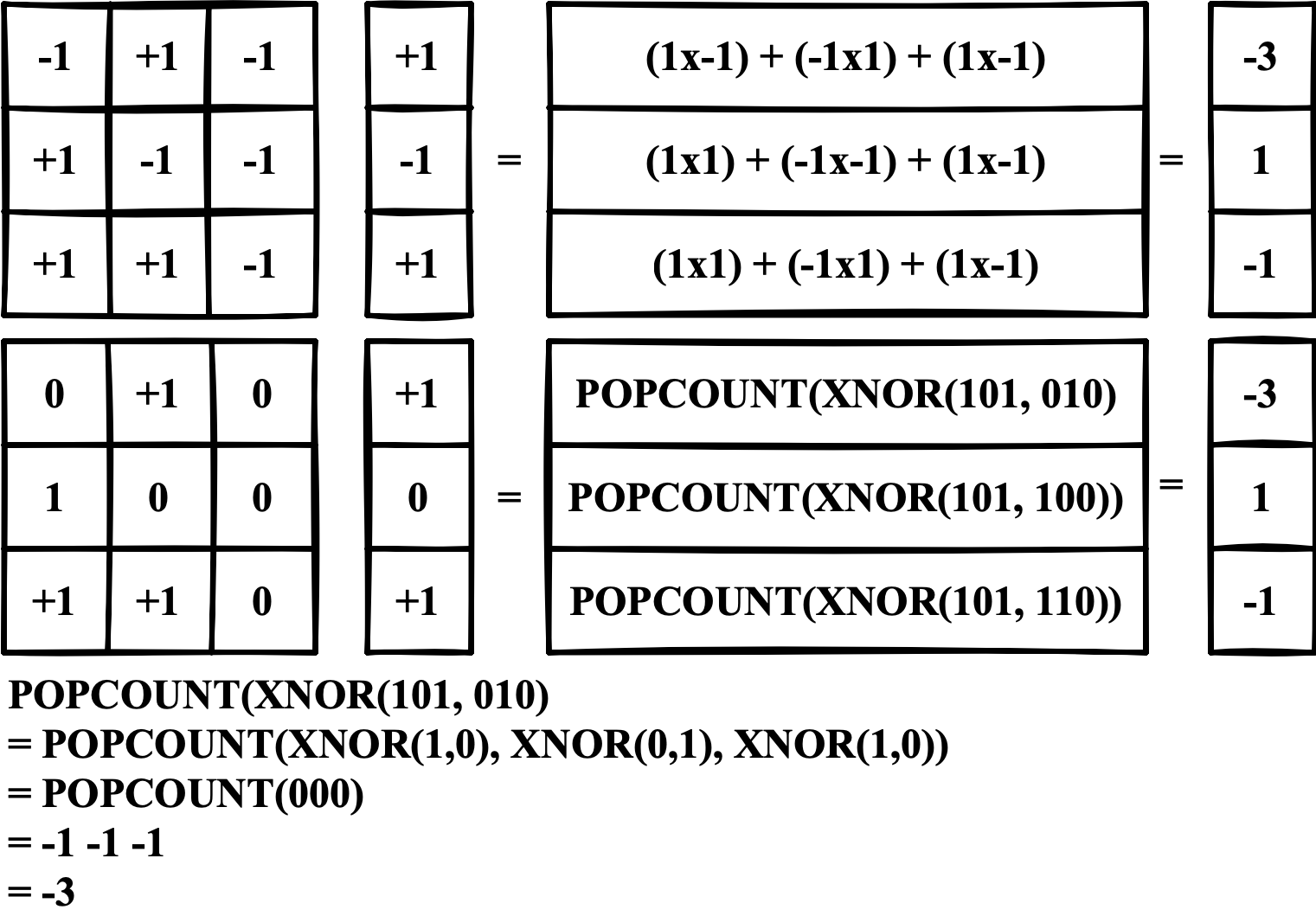}}  
   \caption{XNOR and POPCOUNT in XNOR-Net}
   \label{fig:xnor-popcount}
%   \vspace*{-3mm}
\end{figure}

\noindent 
The convolutions between the matrices (input/activations and weights) are implemented using exclusive-NOR (XNOR) and bit-counting operations. For this, the convolution between two vectors $\in \mathbb{R}^n$ is approximated by the dot products between two vectors $\in \{-1, +1\}^n$ \cite{rastegari2016XNOR}. Thus, convolution between the input $X$ and weight $W$ can be written as: 
\begin{equation}
    Z \approx (sign(X) \odot sign(W)) \odot \alpha\beta
    \label{eqn:sign-dot-conv}
\end{equation}
where $\alpha$ and $\beta$ denote the scaling factors for all the sub tensors in input $X$ and weight $W$ respectively and $\odot$ denotes element-wise multiplication. Due to the binary activations the dot product between $sign(X)$ and $sign(W)$ can be replaced by XNOR and pop-count operations which require extremely little computation. Hence, Equation~\ref{eqn:sign-dot-conv} is reduced to:
\begin{equation}
    Z \approx (X' \circledast W') \odot \alpha\beta
    \label{eqn:xnor-conv}
\end{equation}
where $X' = sign(X)$, $W' = sign(W)$ with all zeros replaced by -1, and $\circledast$ denotes the XNOR and pop-count operations between $X'$ and $W'$. This process is extremely efficient in terms of memory and energy usage. According to Rastegari et al.~ \cite{rastegari2016XNOR} XNOR-Net requires $32x$ less memory and reduces $58x$ computation requirements. Figure~\ref{fig:xnor-popcount} provides an example of how the dot product between $sign(X)$ and $sign(W)$ is replaced by XNOR and pop-count operations between $X'$ and $W'$. 

Like structured pruning based compression works, almost all the works based on XNOR-Net are proposed for computer vision. The current state-of-the-art XNOR models for the benchmark image datasets are: \cite{rastegari2016XNOR} for MNIST, \cite{congXNORNet} for CIFAR-10 and CIFAR-100 and \cite{bulat2020ExpertBinNet} for ImageNet. In contrast, we have found \cite{cerutti2020AudioXnor} to be the only XNOR-Net based work for audio classification. However, that work uses spectrograms as the input to the model which is similar to image classification using XNOR-Net. According to the discussion presented in Section~\ref{sec:handling-audio-data}, using raw audio as input is the appropriate method for audio classification in MCUs. According to our knowledge, there is no such XNOR-Net based work present in the current literature.

\section{Experimental Details}
\label{sec:experiment-details}
All experiments are conducted using Python version 3.7.4, and the GPU versions of Torch 1.8.1. All experimental codes are available at: \href{https://github.com/mohaimenz/pruning-vs-xnor}{https://github.com/mohaimenz/pruning-vs-xnor}.
\subsection{Datasets}
\label{sec:experiment-details-dataset}
The experiments are conducted on three widely used audio benchmark datasets: Environmental Sound, namely ESC-50 and ESC-10 \cite{piczak2015ESC}, and UrbanSound8k~\cite{salamon2014Urbansound8k}. For the image classification experiments, we use the CIFAR-10 and CIFAR-100 benchmark image datasets.  

ESC-50 contains 2000 samples that are equally distributed over 50 disjoint classes. The length of the audio samples is 5s recorded at 16kHz and 44.1kHz. Furthermore, the dataset is partitioned into five folds for cross validation to help researchers obtain directly comparable results. ESC-10 is a subset of ESC-50 with 400 audio samples distributed equally over 10 classes. The subsets of the ESC datasets are as follows:
\\
\begin{math}
    S_{10} = \text{The whole ESC-10 dataset}\\
    S_{20} = S_{10} \cup \{5, 31, 18, 27, 48, 8, 15, 45, 25, 34\}\\
    S_{30} = S_{20} \cup \{3, 14, 23, 36, 43, 7, 22, 28, 30, 49\}\\
    S_{40} = S_{30} \cup \{6, 9, 16, 17, 24, 29, 32, 35, 37, 44\}\\
    S_{50} = \text{The whole ESC-50 dataset}\\
\end{math}

UrbanSound8k contains 8732 labeled audio samples of approx. 4s each, recorded at 22.05kHz. The data are pre-sorted into 10 folds and distributed over 10 classes for easy reproduction and comparison of the performances of different algorithms.

The AudioEvent dataset has 28 classes and 5223 samples unevenly distributed across 28 classes. The data is recorded at 16kHz with a bit depth of 16bit. The subsets of the AudioEvent datasets are as follows:
\\
\begin{math}
    S_{10} = \{0, 2, 5, 9, 11, 12, 17, 21, 25, 26\}\\
    S_{20} = S_{10} \cup \{1, 4, 7, 8, 14, 15, 19, 20, 23, 27\}\\
    S_{28} = \text{The whole AudioEvent dataset}\\
\end{math}

The CIFAR-10 dataset contains 60,000 image samples equally distributed over 10 classes. 50,000 of the samples are used for training and 10,000 for testing. CIFAR-100 has 100 classes, and 60,000 samples are equally distributed across the classes. For each class, there are 500 training samples and 100 test samples. We create the subsets of CIFAR-100 using Equation~\ref{eqn:create-subsets}. The subsets are defined as follows.
\begin{equation}
    S_{x_{i}} = S_{x_{i-1}} \cup \{\frac{x}{10}+m\}
    \label{eqn:cifar-subsets}
\end{equation}
where $x \in \{ 10, 20, \ldots, \lfloor n \rfloor\} \mid n \in [1,100] \quad \text{and} \quad m \in [1,10]$. This gives us the following subsets such as:
\\
\begin{math}
    S_{10} = \text{The whole CIFAR-10 dataset.}\\
    s_{10} = \{0,10,20,30,40,50,60,70,80,90\}\\
    S_{20} = s_{10} \cup\{1,11,21,31,41,51,61,71,81,91\}\\
    S_{30} = S_{20} \cup\{1,11,21,31,41,51,61,71,81,91\}\\
    ...\\
    ...\\
    S_{90} = S_{80} \cup \{9,19,29,39,49,59,69,70,89,99\}\\
    S_{100} = \text{The whole CIFAR-100 dataset.}
\end{math}
\subsection{Data Preprocessing}
For the audio datasets (ESC-10, ..., 50 and UrbanSound8k), we train the DL models with samples of length 30,225, i.e., $\approx 1.51s$ audio at 20kHz (for the AudioEvent dataset the length is 51,215, i.e., $\approx 3.2s$ audio at 16kHz). We use data augmentation as described in \cite{tokozume2017Envnet2} and \cite{mohaimen2021ACDNet} for the audio datasets. For the image datasets, we use random cropping, horizontal flipping, and rotation available in the Transforms module of the PyTorch TorchVision library. 

All implementations of the data augmentation procedures are available in our GitHub repository.  
    
\subsection{Models and Hyperparameters}
The model configuration is available in the GitHub repository, and the hyperparameters for all the experiments conducted on the audio and image datasets are listed in Table~\ref{tab:exp-hyper-params}.
\begin{table}[H]
 \caption{Hyperparameter settings for experiments conducted on Audio Datasets (ESC10, ...,50, UrbanSound8k and AudioEvent) and Image datasets (CIFAR-10,\ldots,100). In the loss functions row, KLD stands for KL Divergence, CE for Cross Entropy. In the Optimizer row, SGD stands for Stochastic Gradient Descent and ADAM for Adaptive Momentum Estimation. In the LR Scheduler row, CosAnLR stands for CosineAnnealingLR.}
%   \begin{center}
    \resizebox{1.0\columnwidth}{!}{%
    \begin{tabular}{|c| c | c | c | c |} 
        \hline
        Datasets$\rightarrow$ &\multicolumn{2}{c|}{Audio Datasets} &\multicolumn{2}{c|}{Image Datasets}\\
      	\cline{2-5}
      	Networks$\rightarrow$     &\multicolumn{2}{c|}{ACDNet, AclNet \& } & \multicolumn{2}{c|} {} \\
      	                        &\multicolumn{2}{c|}{their derivatives}    & \multicolumn{2}{c|} {RESNET-18} \\
      	\cline{2-5}
      	Hyperparams$\downarrow$ & Full Precision & XNOR &   Full Precision    &     XNOR \\
      	\hline
      	Input shape (ch, h, w)  &\multicolumn{2}{c|}{(1, 1, 30225)} & \multicolumn{2}{c|}{(3, 32, 32)} \\
      	\hline
      	Loss function           &\multicolumn{2}{c|}{KLD}   &\multicolumn{2}{c|}{CE}\\
      	\hline
      	Optimizer               &SGD    &ADAM   &SGD    &ADAM   \\
      	\hline
      	Weight decay            &5e-4  & 1e-4   &5e-4   &1e-4   \\
      	\hline
      	Momentum                &0.9   & -      &0.9    &-      \\
      	\hline
      	Initial LR              &0.1   &0.001   &0.1    &0.001\\
      	\hline
      	                  &\multicolumn{2}{c|}{2000 (ESC-10,...50), } &\multicolumn{2}{c|}{}   \\
      	Epochs                       &\multicolumn{2}{c|}{1000 (UrbanSound8k) \&} &\multicolumn{2}{c|}{400}   \\
      	                       &\multicolumn{2}{c|}{1500 (AudioEvent)} &\multicolumn{2}{c|}{}   \\
      	\hline
      	LR Scheduler            & (0.3, 0.6, 0.9) & CosAnLR &\multicolumn{2}{c|}{CosAnLR} \\
      	\hline
      	Warmup epochs           &10 &-  &10 &-  \\
      	\hline
      	LR decay                &$0.1x$ &-  &\multicolumn{2}{c|}{-} \\
      	\hline
      	Batch size              &\multicolumn{4}{c|}{64}      \\
      	\hline
    \end{tabular}}
    \label{tab:exp-hyper-params}
%   \end{center}
\end{table}

\section{Analysis: Pruning vs XNOR-Net}
\label{sec:analysis}
In this section, we compare pruning-and-quantization techniques with XNOR-Net for raw audio classification. Our analysis through extensive experiments on different standard benchmark datasets for audio classification shows that XNOR is more effective for relatively simple problems and very tight constraints on the computational resources; however, the higher classification accuracy achieved by pruning-and-quantization outweighs the benefit of XNOR-Net for problems with complex data characteristics, large number of classes, etc. 

As our starting point for pruning-and-quantization techniques, we use two of the recent state-of-the-art DL networks for raw audio classification called ACDNet \cite{mohaimen2021ACDNet} and AclNet\cite{huang2018Aclnet} as the baseline networks and build their derivatives using pruning-and-quantization as well as the XNOR-Net technique. We test these models on three standard benchmark sound classification datasets - ESC-10 \cite{piczak2015ESC}, ESC-50 \cite{piczak2015ESC}, UrbanSound8k \cite{salamon2014Urbansound8k}, and on subsets of them for an in-depth analysis of the effect of increasing the number of classes. We create subsets of a dataset $S_n$ with $n$ classes as: 
\begin{equation}
    S_x \subseteq S_n \mid x \in \{ 10, 20, \ldots, \lfloor n \rfloor\}
    \label{eqn:create-subsets}
\end{equation}

\noindent
We measure classification accuracy through 5-fold cross validation. The experimental details are provided in Section~\ref{sec:experiment-details}.

\subsection{Pruning-and-Quantization}
\label{sec:analysis-pruning}
We have used the hybrid structured pruning technique proposed in \cite{mohaimen2021ACDNet} to derive Micro-ACDNet by pruning 80\% of the channels from ACDNet. The trained model is first sparsified using the $l0$ norm. It can be expressed as: 
\begin{equation}
    \hat{Z} = W[\chi(W)]
    \label{eqn:hyb-l0-nrom-activation}
\end{equation}
where $W$ denotes the weights of all the layers of the network, and $\chi$ is the function that sorts $W$ and returns the indices of the bottom 95\% of the weights. The channels are then ranked, and the lowest-ranked channel is removed from the network using Equations~\ref{eqn:prune-taylor-criteria},~\ref{eqn:prune-taylor-activation},~\ref{eqn:gen-layer-norm}, and \ref{eqn:gen-rank}. The resulting model using this iterative pruning and fine-tuning process is called Micro-ACDNet~\cite{mohaimen2021ACDNet}. Micro-ACDNet is quantized using a post-training 8-bit quantization technique. We refer to this quantized model as QMicro-ACDNet. The memory and computation requirements of QMicro-ACDNet make it suitable for current off-the-shelf MCUs (see Table~\ref{tab:memory-computation}). We follow the same procedure to derive the respective derivatives of AclNet.

\begin{table}[H]    
    \captionsetup{aboveskip=-2pt}
    \caption{Size and computation requirements for ACDNet, AclNet including their Micro and QMicro versions}
    \centering
    \resizebox{1.0\columnwidth}{!}{%
    \begin{tabular}{| c | c | c | c | c | c | c |} 
    	\hline
      	Models & \multicolumn{2}{c|}{Params (M)} & \multicolumn{2}{c|}{Size (KB)} & \multicolumn{2}{c|}{FLOPs (M)}\\
      	\hline
      	&ACDNet &AclNet &ACDNet &AclNet &ACDNet &AclNet \\
      	\hline
      	Baseline & 4.74 & 10.63 & 18498.00 & 41512.96 & 544.00 & 1806.57\\
      	\hline
      	Micro & 0.13 & 0.13 & 501.00 & 502.00 & 14.82 & 21.50\\
      	\hline
      	QMicro & 0.13 & 0.13 & 133.12 & 128.50 & 14.82 & 21.50\\
      	\hline
    \end{tabular}}
    \label{tab:memory-computation}
\end{table}

Micro-ACDNet and QMicro-ACDNet require $36x$ and $144x$ less memory, respectively, than ACDNet that requires 18.06MB to store its 4.74M parameters. Furthermore, both smaller versions require $37x$ less floating point operations (FLOPs) compared to the base model. For AclNet, the memory reductions for the same derivatives are $86x$ and $324x$ with $84x$ less FLOPs. We note that QMicro version requires the same number of FLOPs as the Micro version but only at quarter precision. If a full 32-bit floating-point unit (FPU) is used, this does not make a practical difference, but it allows us to exploit significant speed gains using 16-bit FPUs, 16-bit FPU modes on full-precision FPUs, and software emulations when no hardware FPU is available.  

\begin{figure*}[ht]
	\centering
    \subfloat[ACDNet]{
        \begin{tikzpicture}
    		\begin{axis}[legend pos=north east,
    			width = 0.8\columnwidth,
    			height = 6cm,
    			xlabel=Classes,
    			grid=major,
    			ylabel=Accuracy(\%),
    			xtick={10,20,30,40,50},
    			ymin=70,ymax=100,
    			ytick={70, 75, 80, 85, 90, 95},
    			legend style={at={(1, 1)},font=\small},
    			legend columns=3,
    			legend style={at={(0.5,-0.25)}, anchor=north,legend cell align=left} %
    			]
    		\addplot[color=black, mark=*] coordinates{
    			(10, 96.75)
    			(20, 92.38)
    			(30, 89.25)
    			(40, 85.94)
    			(50, 87.05)
    		};
    		\addplot[color=blue, mark=*] coordinates{
    			(10, 96.25)
    			(20, 90.13)
    			(30, 86.83)
    			(40, 81.56)
    			(50, 83.25)
    		};
    		\addplot[color=red, mark=*] coordinates{
    			(10, 92.75)
    			(20, 82.55)
    			(30, 81.67)
    			(40, 75.71)
    			(50, 75.50)
    		};
    		\legend{Baseline, Micro, QMicro}
    		\end{axis}
    	\end{tikzpicture}
    % 	\caption{ACDNet}
    	\label{subfig:ACDNet}
    }
    \subfloat[AclNet]{
	\begin{tikzpicture}
		\begin{axis}[legend pos=north east,
			width = 0.8\columnwidth,
			height = 6cm,
			xlabel=Classes,
			grid=major,
			ylabel=Accuracy(\%),
			xtick={10,20,30,40,50},
			ymin=70,ymax=100,
			ytick={70, 75, 80, 85, 90, 95},
			legend style={at={(1, 1)},font=\small},
			legend columns=3,
			legend style={at={(0.5,-0.25)}, anchor=north,legend cell align=left} %
			]
		\addplot[color=black, mark=*] coordinates{
			(10, 95.75)
			(20, 91.12)
			(30, 87.81)
			(40, 85.00)
			(50, 85.70)
		};
		\addplot[color=blue, mark=*] coordinates{
			(10, 94.00)
			(20, 88.50)
			(30, 84.75)
			(40, 79.10)
			(50, 80.05)
		};
		\addplot[color=red, mark=*] coordinates{
			(10, 90.75)
			(20, 80.25)
			(30, 79.34)
			(40, 73.81)
			(50, 72.25)
		};
		\legend{Baseline, Micro, QMicro}
		\end{axis}
	\end{tikzpicture}
	\label{subfig:AclNet}
    }
	\caption{Comparison of prediction accuracy between the baseline models (i.e., ACDNet and AclNet) and the micro versions including their quantized versions (QMicro) on ESC-10,\ldots,50 datasets.}
	\label{plot:esc-accuracy}
% 	\vspace*{-3mm}
\end{figure*}
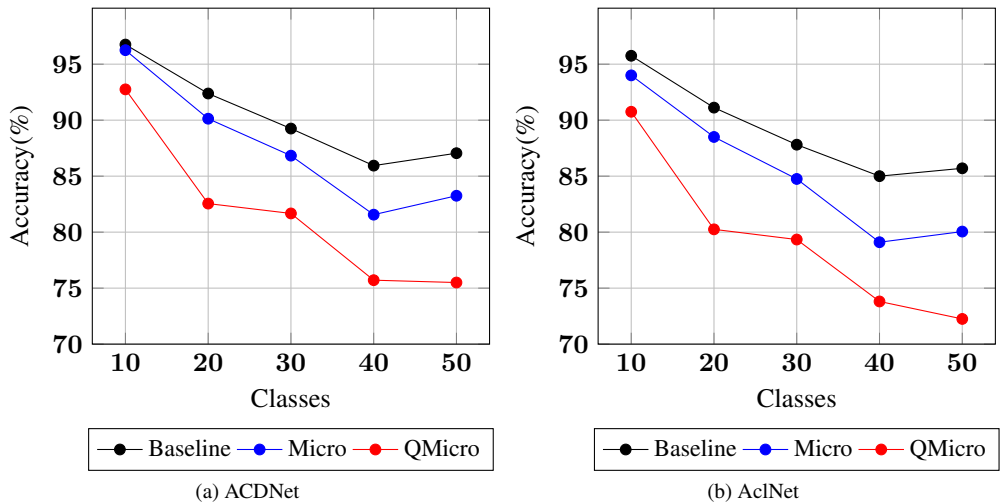

Although the smaller models require fewer resources, they lose classification accuracy because they have less capacity to learn. According to \cite{mohaimen2021ACDNet}, Micro-ACDNet has 80\% less capacity than ACDNet and QMicro-ACDNet is a quarter precision version (8-bit) of Micro-ACDNet. Table~\ref{tab:esc-accuracy} and Figure~\ref{plot:esc-accuracy} present a comparison of the accuracy achieved by the three versions of the both the baseline networks on ESC-10,\ldots,50 datasets (derived using Equation~\ref{eqn:create-subsets}. For further details, see Section~\ref{sec:experiment-details-dataset}). The table and the figure show that all the three versions of both the networks produce state-of-the-art and near state-of-the-art accuracy on ESC-10, however, the accuracy  drops continuously with an increase in the number of classes, as may be expected. 

% \captionsetup[subfloat]{position=top}
\begin{table}[H]
    \captionsetup{aboveskip=-2pt}
    \caption{Prediction accuracy (\%) of Micro versions of ACDNet and AclNet including their quantized versions (QMicro) on ESC-10,\ldots,50 datasets. The column headers '\#Cls' stands for 'No. of Classes' and 'Baseline' stands for the base model (i.e., ACDNet or AclNet).}
    \centering
    \resizebox{1.0\columnwidth}{!}{%
    \begin{tabular}{| c | c | c | c | c | c | c |} 
    	\hline
      	\#Cls. & \multicolumn{3}{c|}{ACDNet} & \multicolumn{3}{c|}{AclNet} \\
      	\hline
      	 & Baseline & Micro & QMicro & Baseline & Micro & QMicro \\
      	\hline
      	10 & 96.75 & 96.25 & 92.75 & 95.75 & 94.00 & 90.75\\
      	\hline
      	20 & 92.38  & 90.13 & 82.55 & 91.12  & 88.50 & 80.25\\
      	\hline
      	30 & 89.25 & 86.83 & 81.67 & 87.81 & 84.75 & 79.34\\
      	\hline
      	40 & 85.94 & 81.56 & 75.71 & 85.00 & 79.10 & 73.81\\
      	\hline
      	50 & 87.05 & 83.25 & 75.50 & 85.70 & 80.05 & 72.25\\
      	\hline
    \end{tabular}}
    \label{tab:esc-accuracy}
\end{table}

The base ACDNet achieves an accuracy of 96.75\% and 87.05\% on ESC-10 and ESC-50, respectively, whereas Micro-ACDNet produces 96.25\% and 83.25\% on ESC-10 and ESC-50, respectively. The quantized version experiences a larger drop in accuracy as the number of classes increases, starting at 92.75\% for ESC-10 and ending with 75.50\% for ESC-50. For AclNet and its different smaller versions, the trend is similar if not the same. Figure~\ref{plot:esc-accuracy} clearly shows these trends. 

\captionsetup[subfloat]{position=bottom}

\begin{table}[H]
    \captionsetup{aboveskip=-2pt}
    \caption{Prediction accuracy (\%) of Micro versions of ACDNet and AclNet including their quantized versions (QMicro) on UrbanSound8k dataset. Here, 'Baseline' stands for the base model (i.e., ACDNet or AclNet)}
    \centering
    \resizebox{0.5\columnwidth}{!}{%
    \begin{tabular}{| c | c | c |} 
    	\hline
      	Models & \multicolumn{2}{c|}{Accuracy (\%)}\\
      	\hline
      	& ACDNet & AclNet \\
      	\hline
      	Baseline & 84.45 & 79.17\\
      	\hline
      	Micro & 78.28 & 75.80\\
      	\hline
      	QMicro & 70.93 & 67.47\\
      	\hline
    \end{tabular}}
    \label{tab:us8k-accuracy}
\end{table}

The scenario is not different when we apply the same networks on the UrbanSound8k dataset. Table~\ref{tab:us8k-accuracy} shows that QMicro-ACDNet has lost almost 13.5\% accuracy compared to the base network. For QMicro-AclNet, the loss in accuracy is 11.7\% compared to the base network. We note that specialized quantization targeted to a particular model can potentially reduce the loss of the accuracy that the models experience during the quantization process. However, improving any technique used to demonstrate the process is beyond the scope of this paper and while a change in quantization techniques may shift the results, the general trends are expected to remain the same.

\subsection{XNOR-Net}
\label{sec:analysis-xnor}
We have applied XNOR-Net technique to derive the XNOR-Net versions of ACDNet (i.e., XACDNet) and AclNet (i.e., XAclNet). Table~\ref{tab:xnor-memory-computation} lists the memory and the computation requirements of the full precision networks and their XNOR counterparts. The table shows that the memory required by the XACDNet is 578KB and XAclNet is 1300KB. These figures are still too much for typical MCUs offering less than 1 MB of RAM, since the parameter spaces alone already leave hardly any room for the actual computations. Hence, we need smaller versions of the original networks whose resource requirements do not exceed the resources available in the MCUs. To compare the network performance, memory, and computation requirements with QMicro-ACDNet, we create Mini-ACDNet such that its XNOR-Net version has similar requirements to QMicro-ACDNet (see Tables \ref{tab:memory-computation} and \ref{tab:xnor-memory-computation}). To derive Mini-ACDNet, we use the same technique as that used to derive Micro-ACDNet, summarized above and fully detailed in~\cite{mohaimen2021ACDNet}. The same procedure is used to derive Mini-AclNet from AclNet.

\begin{table}[H]
    \captionsetup{aboveskip=-2pt}
    \caption{Size and computation requirements for the baseline, Mini and their XNOR-Net versions XBaseline and XMini respectively. Here 'Baseline' stands for the base networks (i.e., ACDNet or AclNet) }
    \centering
    \resizebox{1.0\columnwidth}{!}{%
    \begin{tabular}{| c | c | c | c | c | c | c |} 
    	\hline
      	Model & \multicolumn{2}{c|}{Params (M)} & \multicolumn{2}{c|}{Size (KB)} & \multicolumn{2}{c|}{FLOPs (M)} \\
      	\hline
      	&ACDNet &AclNet &ACDNet &AclNet &ACDNet &AclNet\\
      	\hline
      	 Baseline & 4.74 & 10.63 & 18498 & 41512.96 & 544 & 1806.57\\
      	\hline
      	XBaseline & 4.74 & 10.63 & 578.00 & 1300.48 & 8.88 & 30.02\\
      	\hline
      	Mini & 1.05 & 1.05 & 4096.00 & 4096.00 & 112.00 & 119.00\\
      	\hline
      	XMini & 1.05 & 1.05 & 128.5 & 133.12 & 2.79 & 3.20\\
      	\hline
    \end{tabular}}
    \label{tab:xnor-memory-computation}
    % \vspace*{-3mm}
\end{table}

The sizes of the XNOR-Net versions of ACDNet and AclNet are calculated according to \cite{rastegari2016XNOR}. The number of binary operations required for the networks is calculated according to \cite{liu2018BiRealNet}. We express the computation (Binary operation + FLOPS) required for an XNOR network as FLOPs for simplicity. If $a$ and $b$ are the FLOPs required for the first and the last full precision layers of the XNOR network and $x$ is the total number of FLOPs of the full precision version of the network, then the calculation of FLOPs of an XNOR network can be expressed as $FLOPs = a + (x-(a+b))/64 + b$.

Table~\ref{tab:xnor-memory-computation} also shows that XMini versions are extremely small (128.5KB and 133.12KB) and require considerably less computation. This clearly allows the network to be deployed on the current off-the-shelf MCUs. Furthermore, from Table~\ref{tab:xnor-esc-accuracy}, we can see that the XNOR networks produce reasonable accuracy for a smaller number of classes (ESC-10 with 10 classes). However, as  the number of classes increases, the network performance of the XNOR versions decreases rapidly.

% \captionsetup[subfloat]{position=top}
\begin{table}[H]
    \captionsetup{aboveskip=-2pt}
    \caption{Prediction accuracy (\%) of Baselines and Mini versions including their XNOR-Net versions on ESC-10,...,50 datasets. The column header '\#Cls' stands for 'No. of Classes' and 'Baseline' stands for the base network (i.e., ACDNet or AclNet)}
    \centering
    \resizebox{1.0\columnwidth}{!}{%
    \begin{tabular}{| c | c | c | c | c | c | c | c | c |} 
    	\hline
      	\#Cls. & \multicolumn{4}{c|}{ACDNet} & \multicolumn{4}{c|}{AclNet}\\
      	\hline
      	 & Baseline & XBaseline & Mini & XMini & Base & XBase & Mini & XMini\\
      	\hline
      	10 & 96.75 & 91.25 &  96.75 & 82.25 & 95.75 & 89.70 & 93.50 & 80.50\\
      	\hline
      	20 & 92.38 & 77.25 & 91.12 & 54.75 & 91.12 & 66.00 & 88.38 & 52.75\\
      	\hline
      	30 & 89.25 & 70.42 & 88.58 & 46.83 & 87.81 & 57.33 & 85.6 & 48.00\\
      	\hline
      	40 & 85.94 & 61.87 & 84.50 & 38.56 & 85.00 & 48.00 & 81.44 & 35.60\\
      	\hline
      	50 & 87.05 & 56.40 & 85.60 & 31.70 & 85.70 & 39.85 & 82.80 & 31.55\\
      	\hline
    \end{tabular}}
    \label{tab:xnor-esc-accuracy}
    % \vspace*{-3mm}
\end{table}

\begin{figure*}[htb]
	\centering
	\subfloat[ACDNet]{
	\begin{tikzpicture}
		\begin{axis}[
			width = 0.8\columnwidth,
			xlabel=Classes,
			grid=major,
			ylabel=Accuracy (\%),
			xtick={10,20,30,40,50},
			ymin=20,ymax=100,
			ytick={20,30,40,50,60,70,80,90,100},
			legend columns=3,
			legend style={at={(0.5,-0.25)}, anchor=north,legend cell align=left} %
			]
		%Full precision ACDNet
		\addplot[color=black, mark=*] coordinates{
			(10, 96.75)
			(20, 92.38)
			(30, 89.25)
			(40, 85.94)
			(50, 87.05)
		};
		%Full precision Mini-ACDNet
		\addplot[color=blue!80, mark=*] coordinates{
			(10, 96.75)
			(20, 91.12)
			(30, 88.58)
			(40, 84.50)
			(50, 85.60)
		};
		%Full precision Micro-ACDNet
		\addplot[color=gray, mark=*] coordinates{
			(10, 96.25)
			(20, 90.13)
			(30, 86.83)
			(40, 81.56)
			(50, 83.25)
		};
		%XACDNet
		\addplot[color=red!50, mark=*] coordinates{
			(10, 91.25)
			(20, 77.25)
			(30, 70.42)
			(40, 61.87)
			(50, 56.40)
		};
		%XMini-ACDNet
		\addplot[color=purple!30, mark=*] coordinates{
			(10, 82.25)
			(20, 54.75)
			(30, 46.83)
			(40, 38.56)
			(50, 31.70)
		};
		%QMicro-ACDNet
		\addplot[color=purple, mark=*] coordinates{
			(10, 92.75)
			(20, 82.55)
			(30, 81.67)
			(40, 75.71)
			(50, 75.50)
		};
		\legend{Baseline, Mini, Micro, XBaseline, XMini, QMicro}
		\end{axis}
	\end{tikzpicture}
	\label{plot:acdnet-xnor-esc-accuracy}
	}
	\subfloat[AclNet]{
	\begin{tikzpicture}
		\begin{axis}[
			width = 0.8\columnwidth,
			xlabel=Classes,
			grid=major,
			ylabel=Accuracy (\%),
			xtick={10,20,30,40,50},
			ymin=20,ymax=100,
			ytick={20,30,40,50,60,70,80,90,100},
			legend columns=3,
			legend style={at={(0.5,-0.25)}, anchor=north,legend cell align=left} %
			]
		%Full precision AclNet
		\addplot[color=black, mark=*] coordinates{
			(10, 95.75)
			(20, 91.12)
			(30, 87.81)
			(40, 85.00)
			(50, 85.70)
		};
		%Full precision Mini-AclNet
		\addplot[color=blue!80, mark=*] coordinates{
			(10, 93.50)
			(20, 88.38)
			(30, 85.60)
			(40, 81.44)
			(50, 82.80)
		};
		%Full precision Micro-AclNet
		\addplot[color=gray, mark=*] coordinates{
			(10, 94.00)
			(20, 88.50)
			(30, 84.75)
			(40, 79.10)
			(50, 80.05)
		};
		%XAclNet
		\addplot[color=red!50, mark=*] coordinates{
			(10, 89.70)
			(20, 66.00)
			(30, 57.33)
			(40, 48.00)
			(50, 39.85)
		};
		%XMini-AclNet
		\addplot[color=purple!30, mark=*] coordinates{
			(10, 80.50)
			(20, 52.75)
			(30, 48.00)
			(40, 35.60)
			(50, 31.55)
		};
		%QMicro-AclNet
		\addplot[color=purple, mark=*] coordinates{
			(10, 90.75)
			(20, 80.25)
			(30, 79.34)
			(40, 73.81)
			(50, 72.25)
		};
		\legend{Baseline, Mini, Micro, XBaseline, XMini, QMicro}
		\end{axis}
	\end{tikzpicture}
	\label{plot:aclnet-xnor-esc-accuracy}
	}
	\caption{Comparison of prediction accuracy between Baseline, Mini including their XNOR-Net versions and Micro including its quantized version (QMicro) on ESC-10,\ldots,50 datasets. The Baseline network is AcdNet or AclNet. XMini and QMicro versions have approximately the same memory requirements.}
	\label{plot:xnor-esc-accuracy}
% 	\vspace*{-3mm}
\end{figure*}
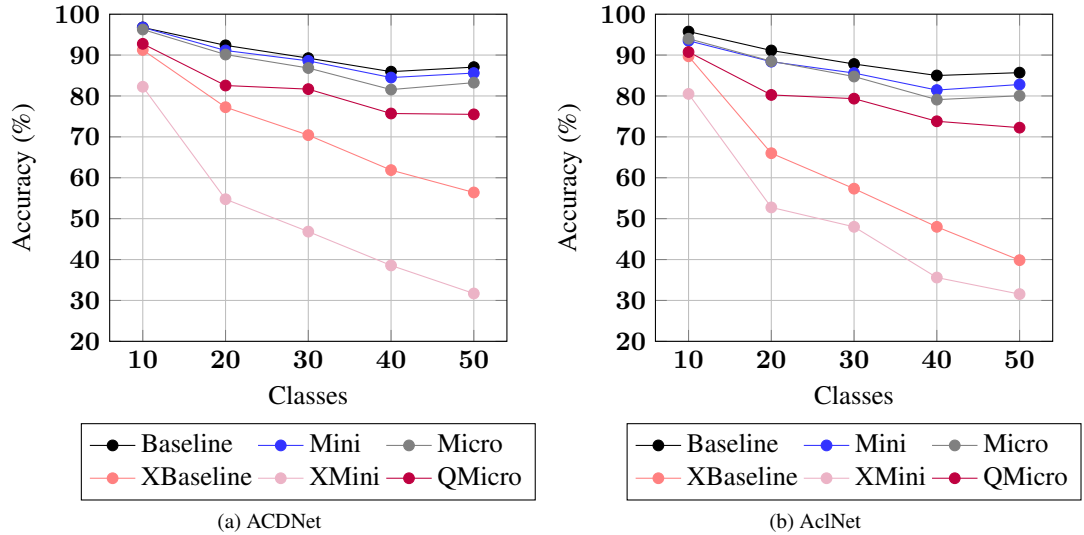

Figure~\ref{plot:xnor-esc-accuracy} shows the performance graph of the Baseline models (i.e., ACDNet and AclNet) and their mini, pruning-and-quantization and XNOR counterparts. The line at the bottom (XMini) shows how extremely the smallest XNOR-Net is affected when the number of classes increases, while QMicro manages degrade much more gracefully. For all the XNOR-Net networks (XBaseline and XMini versions of the Baselines), the slope is much steeper than that for the other compression methods.

\captionsetup[subfloat]{position=bottom}

\captionsetup[subfloat]{position=bottom}
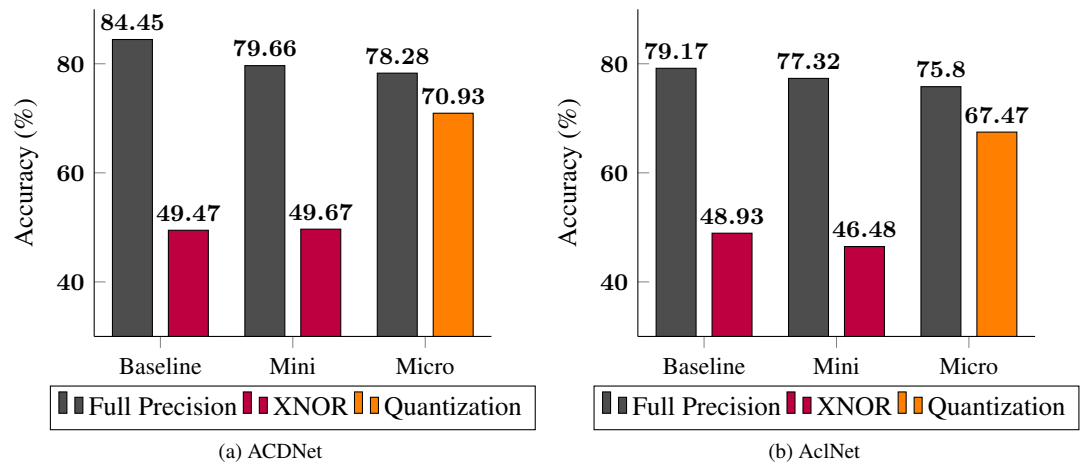
\begin{figure*}[htb]
	\centering
	\subfloat[ACDNet]{
	\begin{tikzpicture}
		\begin{axis}[
			ybar=5pt,
   			width = 0.8\columnwidth, 
   			symbolic x coords={Baseline, Mini, Micro}, 
   			xtick=data,
   			ymin=30, ymax=90,
   			ylabel = Accuracy (\%),
   			bar width=15pt,
			axis lines*=left,
   			ticklabel style = {font=\small},
   			nodes near coords,
   			every node near coord/.append style={font=\small},
  			enlarge x limits=0.25,
  			legend columns=-1,
   			legend style={at={(0.5,-0.15)}, anchor=north,legend cell align=left} %
   			]
    		\addplot[plot 0, bar group size={0}{2}] coordinates {
       			(Baseline, 84.45)
       			(Mini, 79.66)
       			(Micro, 78.28)
    		};
    		\addplot[plot 1, bar group size={1}{2}] coordinates {
        		(Baseline, 49.47)
        		(Mini, 49.67)
    		};
    		\addplot [plot 2, bar group size={1}{2}] coordinates {
      			(Micro, 70.93)
    		};
    		\legend{Full Precision, XNOR, Quantization}
		\end{axis}
	\end{tikzpicture}
	\label{subplot:acdnet-xnor-us8k-accuracy}
	}
	\subfloat[AclNet]{
	\begin{tikzpicture}
		\begin{axis}[
			ybar=5pt,
   			width = 0.8\columnwidth, 
   			symbolic x coords={Baseline, Mini, Micro}, 
   			xtick=data,
   			ymin=30, ymax=90,
   			ylabel = Accuracy (\%),
   			bar width=15pt,
			axis lines*=left,
   			ticklabel style = {font=\small},
   			nodes near coords,
   			every node near coord/.append style={font=\small},
  			enlarge x limits=0.25,
  			legend columns=-1,
   			legend style={at={(0.5,-0.15)}, anchor=north,legend cell align=left} %
   			]
    		\addplot[plot 0, bar group size={0}{2}] coordinates {
       			(Baseline, 79.17)
       			(Mini, 77.32)
       			(Micro, 75.80)
    		};
    		\addplot[plot 1, bar group size={1}{2}] coordinates {
        		(Baseline, 48.93)
        		(Mini, 46.48)
    		};
    		\addplot [plot 2, bar group size={1}{2}] coordinates {
      			(Micro, 67.47)
    		};
    		\legend{Full Precision, XNOR, Quantization}
		\end{axis}
	\end{tikzpicture}
	\label{subplot:aclnet-xnor-us8k-accuracy}
	}
	\caption{Comparison of prediction accuracy between Baseline, Mini including their XNOR-Net versions and Micro including its quantized version (QMicro) on the UrbanSound8k dataset. The Baseline network is ACDNet or AclNet. XMini and QMicro versions have approximately the same memory requirements.}
	\label{plot:xnor-us8k-accuracy}
% 	\vspace*{-3mm}
\end{figure*}
Figure~\ref{plot:xnor-us8k-accuracy} shows the performance of the same networks on the UrbanSound8k dataset, another widely used audio benchmark. From the graph, we observe that the results for UrbanSound8k confirm our findings for ESC-10,\ldots,50.

In summary, we observe that the memory requirements of QMicro and XMini versions are essentially the same. XMini-ACDNet requires $7.97x$ less FLOPs than QMicro-ACDNet, on the other hand, XMini-AclNet requires $6.72x$ less FLOPs that QMicro-AclNet. However, the XNOR-based models do not reach comparable classification accuracies when the number of classes is larger. Although the pruning-and-quantization-based QMicro-ACDNet also sees a loss in accuracy, the accuracy is still reasonable for larger problems given its minuscule model size. This observation is consistent with both the baseline models and their derivatives.

\subsection{Extended Analysis}
\label{sec:analysis-other}
\subsubsection{Comparing with Existing Work}
To the best of our knowledge, Cerutti et al.~\cite{cerutti2020AudioXnor} have presented the only XNOR network for audio classification so far, termed BNN-GAP8. They have used a different benchmark, namely, the less commonly used  AudioEvent dataset~\cite{takahashi2016DNNAndDataAugmentationAcoustic}. We cannot test BNN-GAP8 on the most widely used benchmarks, ESC50 and Urbansound-8k, since  BNN-GAP8 has not been described in sufficient detail in~\cite{cerutti2020AudioXnor} to make it reproducible.

\begin{figure*}[htb]
	\centering
	\subfloat[ACDNet]{
	\begin{tikzpicture}
		\begin{axis}[
			width = 0.8\columnwidth,
			xlabel=Classes,
			grid=major,
			ylabel=Accuracy (\%),
			xtick={10,20,28},
			ymin=40,ymax=100,
			ytick={40,50,60,70,80,90,100},
			legend columns=3,
			legend style={at={(0.5,-0.25)}, anchor=north,legend cell align=left} %
			]
		% ACDNet
		\addplot[color=black, mark=*] coordinates{
			(10, 96.25)
			(20, 92.82)
			(28, 92.57)
		};
		% XACDNet
		\addplot[color=red!65, mark=*] coordinates{
			(10, 78.5)
			(20, 53.27)
			(28, 49.82)
		};
		% Mini-ACDNet
		\addplot[color=green!80, mark=x] coordinates{
			(10, 95.86)
			(20, 92.93)
			(28, 92.49)
		};
		% XMini-ACDNet
		\addplot[color=purple!40, mark=*] coordinates{
			(10, 74.16)
			(20, 53.48)
			(28, 43.4)
		};
		% Micro-ACDNet
		\addplot[color=blue, mark=*] coordinates{
			(10, 95.66)
			(20, 90.03)
			(28, 89.69)
		};
		% QMicro-ACDNet
		\addplot[color=purple, mark=*] coordinates{
			(10, 94.08)
			(20, 86.71)
			(28, 84.98)
		};
		\legend{Baseline, XBaseline, Mini, XMini, Micro, QMicro}
		\end{axis}
	\end{tikzpicture}
	}
	\subfloat[AclNet]{
	\begin{tikzpicture}
		\begin{axis}[
			width = 0.8\columnwidth,
			xlabel=Classes,
			grid=major,
			ylabel=Accuracy (\%),
			xtick={10,20,28},
			ymin=40,ymax=100,
			ytick={40,50,60,70,80,90,100},
			legend columns=3,
			legend style={at={(0.5,-0.25)}, anchor=north,legend cell align=left} %
			]
		% ACDNet
		\addplot[color=black, mark=*] coordinates{
			(10, 96.25)
			(20, 92.82)
			(28, 92.57)
		};
		% XACDNet
		\addplot[color=red!65, mark=*] coordinates{
			(10, 78.5)
			(20, 53.27)
			(28, 49.82)
		};
		% Mini-ACDNet
		\addplot[color=green!80, mark=x] coordinates{
			(10, 95.86)
			(20, 92.93)
			(28, 92.49)
		};
		% XMini-ACDNet
		\addplot[color=purple!40, mark=*] coordinates{
			(10, 74.16)
			(20, 53.48)
			(28, 43.4)
		};
		% Micro-ACDNet
		\addplot[color=blue, mark=*] coordinates{
			(10, 95.66)
			(20, 90.03)
			(28, 89.69)
		};
		% QMicro-ACDNet
		\addplot[color=purple, mark=*] coordinates{
			(10, 94.08)
			(20, 86.71)
			(28, 84.98)
		};
		\legend{Baseline, XBaseline, Mini, XMini, Micro, QMicro}
		\end{axis}
	\end{tikzpicture}
	}
	\caption{Comparison of prediction accuracy between the Baseline, Mini including their XNOR-Net versions and Micro including its quantized version (QMicro) on AudioEvent datasets and its subsets (10, 20 and 28 classes). The Baseline model is ACDNet or AclNet. XMini and QMicro versions have approximately the same memory requirements.}
	\label{plot:ae-accuracy}
% 	\vspace*{-3mm}
\end{figure*}
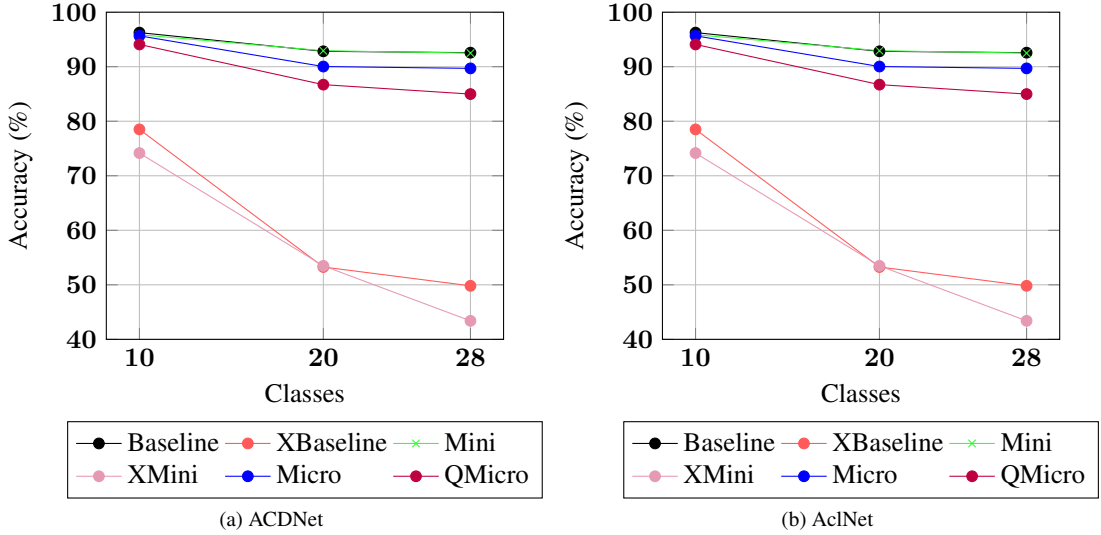

To facilitate a direct comparison, we extend our analysis to this dataset, which has also been used in a several other studies~\cite{takahashi2016DNNAndDataAugmentationAcoustic, meyer2017CNNForAED}. These results are consistent with what we have seen above for the most widely used standard benchmarks. Table~\ref{tab:ae-accuracy} and Figure~\ref{plot:ae-accuracy} show the performance of our base nets on the AudioEvent dataset and its smaller subsets.

\captionsetup[subfloat]{position=bottom}

\begin{table}[H]
    \captionsetup{aboveskip=-2pt}
    \caption{Prediction accuracy (\%) of the Baseline, Mini including their XNOR-Net versions and Micro including its quantized version (QMicro) on AudioEvent-10,20,28 datasets. The column header '\#Cls' stands for 'No. of Classes' and the Baseline model is ACDNet or AclNet. XMini and QMicro versions have approximately the same memory requirements.}
    \centering
    \resizebox{1.0\columnwidth}{!}{%
    \begin{tabular}{| c | c | c | c | c | c | c |} 
        \hline
        \#Cls.$\rightarrow$ & \multicolumn{2}{c|}{10} & \multicolumn{2}{c|}{20} & \multicolumn{2}{c|}{28}\\
        \hline
        Models$\downarrow$ & ACDNet & AclNet & ACDNet & AclNet & ACDNet & AclNet\\
        \hline
        Baseline & 96.25 & 94.05 & 92.82 & 90.46 & 92.57 & 90.15 \\
        \hline
        XBaseline & 78.05 & 77.08 & 53.27 & 50.84 & 49.83 & 41.34 \\
        \hline
        Mini & 95.86 & 93.45 & 92.93 & 90.82 & 92.49 & 89.99 \\
        \hline
        XMini & 74.16 & 73.33 & 53.48 & 50.09 & 43.40 & 44.61 \\
        \hline
        Micro & 95.66 & 94.06 & 90.03 & 88.35 & 89.69 & 87.51 \\
        \hline
        QMicro & 94.08 & 92.48 & 86.71 & 84.28 & 84.98 & 81.57 \\
        \hline
    \end{tabular}}
    \label{tab:ae-accuracy}
\end{table}

However, the larger resource requirements of our base networks do not allow us a direct and fair comparison to the BNN-GAP8 network presented in~\cite{cerutti2020AudioXnor}. Therefore, we derive two additional smaller networks (QNano-ACDNet and XGAP8-ACDNet) with memory requirements equivalent to BNN-GAP8. QNano-ACDNet is an 8-bit quantized version of the associated full-precision network Nano-ACDNet that is derived by pruning the channels from ACDNet. XGAP8-ACDNet is an XNOR network derived from the full precision network GAP8-ACDNet. GAP8-ACDNet is also derived by pruning channels from ACDNet. The equivalent derivatives are also derived from AclNet. The new models are constructed using the same procedures as described above. The tests of these networks on the AudioEvent dataset used in~\cite{cerutti2020AudioXnor} are summarized in Table~\ref{tab:ae-gap8-comparison}. 

% \captionsetup[subfloat]{position=top}
\begin{table}[H]
    \captionsetup{aboveskip=-2pt}
    \caption{Prediction accuracy (\%) of BNN-GAP8~\cite{cerutti2020AudioXnor},  Nano and GAP8 versions of both the baselines including the XNOR-Net for GAP8 (i.e., XGAP8) and the quantized version of Nano (i.e., QNano) on AudioEvent (28 classes) dataset}
    \centering
    \resizebox{1.0\columnwidth}{!}{%
    \begin{tabular}{| c | c | c | c | c | c | c |} 
    	\hline
      	Models & \multicolumn{2}{c|}{Accuracy (\%)} & \multicolumn{2}{c|}{Memory (KB)} & \multicolumn{2}{c|}{FLOPs (M)} \\
      	\hline
      	CNN-CNP \cite{meyer2017CNNForAED} & \multicolumn{2}{c|}{85.1} & \multicolumn{2}{c|}{1766} & \multicolumn{2}{c|}{2478} \\
      	\hline
      	BNN-GAP8 \cite{cerutti2020AudioXnor} & \multicolumn{2}{c|}{77.9} & \multicolumn{2}{c|}{58} & \multicolumn{2}{c|}{1768} \\
      	\hline
      	& ACDNet & AclNet & ACDNet & AclNet & ACDNet & AclNet \\
      	\hline
      	Nano & 86.22 & 84.75 & 245 & 245 & 10.54 & 11.37\\
      	\hline
      	QNano & 82.34 & 80.15 & 61 & 61 & 10.54 & 11.37\\
      	\hline
      	GAP8 & 90.35 & 87.57 & 1960 & 1960 & 40.51 & 54.02\\
      	\hline
      	XGAP8 & 35.48 & 35.93 & 61 & 61 & 1.60 & 2.18\\
      	\hline
    \end{tabular}}
    \label{tab:ae-gap8-comparison}
\end{table}

BNN-GAP8 shows good performance but the quantized networks QNano-ACDNet and QNano-AclNet clearly outperforms it for equivalent storage requirements. The computational requirements for BNN-GAP8 are given in MACs rather than FLOPs in~\cite{cerutti2020AudioXnor}. Using the same re-scaling as detailed above, we arrive at 1768M FLOPs for BNN-GAP8. The quantized  QNano networks compare very favorably, using only 10.54M FLOPs for the ACDNet version and 11.37M FLOPs for the AclNet version. In a nutshell, the pruned-and-quantized QNano versions deliver better performance than BNN-GAP8 XNOR-Net  with smaller resource requirements.

The performance of BNN-GAP8 on the 28-class problem is much better than that of the the XNOR networks derived from ACDNet and AclNet (i.e., XGAP8-ACDNet and XGAP8-AclNet). The results for the latter are consistent with the performance drop established for the slightly larger XMini-ACDNet on ESC30 (ESC50 reduced to 30 classes, cf.\ Table~\ref{tab:xnor-esc-accuracy}).

How can BNN-GAP8 achieve such good performance on a 28-class problem? 
The likely reason is that BNN-GAP8 benefits from using spectrograms as input to the network. This means that in the BNN-GAP8 implementation much of the necessary full precision computation required, for which a binary net is not suitable, is encapsulated outside of the network in the conversion of raw audio to spectrograms. The other XNOR networks in this comparison are deprived of this possibility because they perform end-to-end classification for raw audio input and thus must perform all required computations within the network.

\captionsetup[subfloat]{position=bottom}
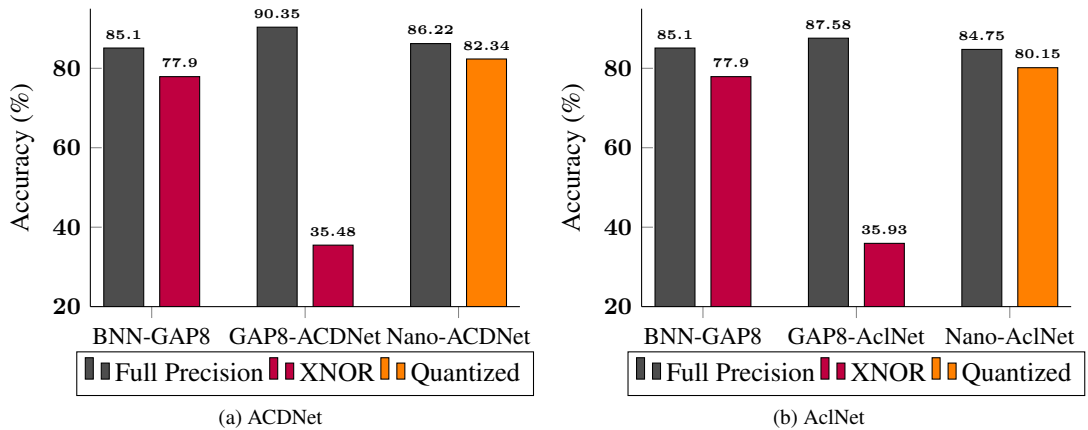
\begin{figure*}[htb]
	\centering
	\subfloat[ACDNet]{
	\begin{tikzpicture}
		\begin{axis}[
  			width = 0.85\columnwidth, 
      	    height = 5.5cm,
  			ybar = 1pt,
  			bar width = 15pt,
  			symbolic x coords={BNN-GAP8, GAP8-ACDNet, Nano-ACDNet}, 
  			xtick=data,
  			ymin=20, ymax=95,
  			ylabel = Accuracy (\%),
			axis lines*=left,
  			ticklabel style = {font=\small},
  			nodes near coords,
  			every node near coord/.append style={font=\tiny},
  			enlarge x limits=0.2,
  			legend columns=-1,
  			legend style={at={(0.5,-0.15)}, anchor=north,legend cell align=left} %
  			]
    		\addplot[plot 0, bar group size={0}{2}] coordinates {
       			(BNN-GAP8, 85.1)
       			(GAP8-ACDNet, 90.35)
       			(Nano-ACDNet, 86.22)
    		};
    		\addplot[plot 1, bar group size={1}{2}] coordinates {
        		(BNN-GAP8, 77.9)
        		(GAP8-ACDNet, 35.48)
    		};
    		\addplot [plot 2, bar group size={1}{2}] coordinates {
      			(Nano-ACDNet, 82.34)
    		};
    		\legend{Full Precision, XNOR, Quantized}
		\end{axis}
	\end{tikzpicture}
	\label{plot:ae-acdnet-gap8-comparison}
	}
	\subfloat[AclNet]{
    \begin{tikzpicture}
	\begin{axis}[
  		width = 0.85\columnwidth,
  		height = 5.5cm,
  		ybar = 1pt,
  		bar width = 15pt,
  		symbolic x coords={BNN-GAP8, GAP8-AclNet, Nano-AclNet}, 
  		xtick=data,
  		ymin=20, ymax=95,
  		ylabel = Accuracy (\%),
		axis lines*=left,
  		ticklabel style = {font=\small},
  		nodes near coords,
  		every node near coord/.append style={font=\tiny},
  		enlarge x limits=0.2,
  		legend columns=-1,
  		legend style={at={(0.5,-0.15)}, anchor=north,legend cell align=left} %
  		]
		\addplot[plot 0, bar group size={0}{2}] coordinates {
   			(BNN-GAP8, 85.1)
   			(GAP8-AclNet, 87.58)
   			(Nano-AclNet, 84.75)
		};
		\addplot[plot 1, bar group size={1}{2}] coordinates {
    		(BNN-GAP8, 77.9)
    		(GAP8-AclNet, 35.93)
		};
		\addplot [plot 2, bar group size={1}{2}] coordinates {
  			(Nano-AclNet, 80.15)
		};
		\legend{Full Precision, XNOR, Quantized}
	\end{axis}
	\end{tikzpicture}
	\label{plot:ae-aclnet-gap8-comparison}
	}
	\caption{Comparison of prediction accuracy between BNN-GAP8 \cite{cerutti2020AudioXnor}, GAP8-ACDNet and Nano-ACDNet on AudioEvent dataset for 28 classes. BNN-GAP8 runs on spectrogram, in contrast, GAP8-ACDNet and Nano-ACDNet run on raw audio. GAP8-ACDNet has approximately the same resource requirements of BNN-GAP8. QNano-ACDNet has equivalent memory requirement of the XNOR versions of BNN-GAP8 and XGAP8-ACDNet}
	\label{plot:ae-gap8-comparison}
% 	\vspace*{-1mm}
\end{figure*}

From a pragmatic perspective, using spectrograms may be a useful approach for handling simple audio classification problems on MCUs that feature a DSP with hardware support for FFT computation, which is the basis of generating spectrograms. However, as has been shown in Table~\ref{tab:ae-gap8-comparison} and Figure~\ref{plot:ae-gap8-comparison}, even where such support exists, forgoing it and using end-to-end classification with a pruned-and-quantized network instead still offers performance and resource benefits over the XNOR network. Furthermore, from the broader literature on DNN audio classification, it is evident that  purely spectrogram-based classification does not always allow us to achieve state-of-the-art performance with conventional CNN architectures and that features learned from raw audio or multi-channel features are preferable for more difficult benchmarks~\cite{kim2020MultiChannel, nanni2021EnsenbleCNNAudio, kumar2020SeqSelfTeaching, gong2021AudioSpectroTransformer, sailor2017UnSupervisedFilterbank}.

% \subsubsection{Analysis with a Second Network}
% \input{sections/subsections/analysis-aclnet}

\subsubsection{Analysis with Image Datasets}
Most of the work on XNOR nets so far has taken place in the image domain. A comparison with this body of work is instructive. We summarize the state-of-the-art for the most widely used datasets in Table~\ref{tab:sota-image}.  The accuracy achieved by XNOR nets  is impressive (second last column), but it has to be noted that these nets are significantly larger than our target size. None of these models fit on the relevant MCUs with the single exception of the one for MNIST. This is a very simple dataset with only 10 classes. 

\begin{table}[H]
    \captionsetup{aboveskip=-2pt}
    \caption{State-of-the-art accuracy (\%) for various image datasets. '\#Cls' stands for 'number of Classes'.}
    % \centering
    \resizebox{1.0\columnwidth}{!}{%
    \begin{tabular}{| c | c | c | c | c | } 
    	\hline
      	Datasets (\#cls.) & \multicolumn{2}{c|}{Full Precision} & \multicolumn{2}{c|}{XNOR} \\
      	\hline
      	 &Accuracy & Size (MB) & Accuracy & Size (MB)\\
      	\hline
      	MNIST (10) & 99.87 \ignore{Byerly et al. }\cite{Byerly2021NoRouting} & 5.8 & 99.23 \ignore{Rastegari et al. }\cite{rastegari2016XNOR} & 0.10 \\
      	\hline
      	CIFAR-10 (10) & 99.50 \ignore{Dosovitskiy et al. }\cite{dosovitskiy2020image} & 2411 & 88.74 \ignore{Cong }\cite{congXNORNet} & 1.3 \\
      	\hline
      	CIFAR-100 (100) & 96.08 \ignore{Tan }\cite{tan2019efficientnet} & - & 77.80 \ignore{Bulat et al. }\cite{bulat2020ExpertBinNet} & 7.8 \\
      	\hline
      	Imagenet (1000) & 90.45 \ignore{Zhai et al. }\cite{zhai2021scaling} & 7030 & 71.20 \ignore{Bulat et al. }\cite{bulat2020ExpertBinNet} & 7.8 \\
      	\hline
    \end{tabular}}
    \label{tab:sota-image}
%   \vspace*{-5mm}
\end{table}

\begin{figure}[H]
	\centering
	\begin{tikzpicture}
		\begin{axis}[legend pos= north east,
			width = 0.8\columnwidth,
			height = 4.8cm,
			xlabel=Classes,
			grid=major,
			ylabel=Accuracy(\%),
			xtick={10,20,30,40,50,60,70,80,90,100},
			ymin=40,ymax=100,
			ytick={50,60,70,80,90,100},
			legend style={at={(1, 1)},font=\small},
			legend columns=1
			]
		\addplot[color=blue, mark=*] coordinates{
			(10, 93.29)
			(20, 85.70)
			(30, 80.87)
			(40, 76.67)
			(50, 75.18)
			(60, 72.28)
			(70, 72.70)
			(80, 71.81)
			(90, 71.61)
			(100, 71.49)
		};
		\addplot[color=red, mark=*] coordinates{
			(10, 80.24)
			(20, 69.45)
			(30, 60.90)
			(40, 57.00)
			(50, 54.36)
			(60, 49.77)
			(70, 51.20)
			(80, 49.51)
			(90, 50.22)
			(100, 49.56)
		};
		\legend{Full Precision, XNOR}
		\end{axis}
	\end{tikzpicture}
	\caption{Prediction accuracy of RESNET-18 and its XNOR-Net version on CIFAR-10,20,30,....,100 datasets}
	\label{plot: resnet-18-cifar}
% 	\vspace*{-1mm}
\end{figure}
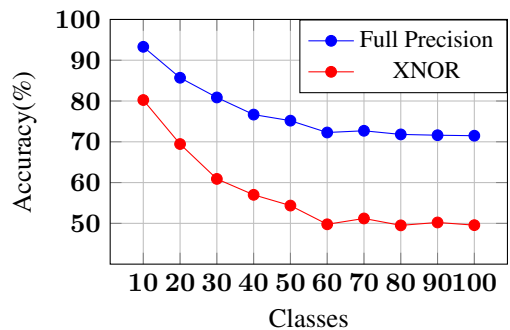

To verify whether our own results for audio classification are consistent with the image domain, we  conducted additional experiments on image classification using XNOR. We have used the XNOR version of RESNET-18~\cite{he2016ResNet} to classify the widely used benchmark image datasets CIFAR-10 and CIFAR-100. We have created variably sized subsets of CIFAR-100 to determine whether the trend of the loss in accuracy is similar to audio classification. The experimental details are provided in Section~\ref{sec:experiment-details}. Table~\ref{tab:resnet18-cifar} and Figure~\ref{plot: resnet-18-cifar} summarize the results. The sizes of the full precision models  are between 42.63MB and 42.80MB and the computation requirements between 95.17M FLOPs and 95.21M FLOPs. For the XNOR-Net version, the model size is approximately 1.34MB using 2.06M FLOPs. The results of this experiment are consistent with those for the audio domain, and it is clearly visible that a similar performance drop for XNOR occurs as the number of classes increases.

\begin{table}[H]
    \captionsetup{aboveskip=-1pt}
    \caption{RESNET-18 and its XNOR-Net version on CIFAR-10,20,30,....,100 datasets.}
    \centering
    \resizebox{0.60\columnwidth}{!}{%
    \begin{tabular}{| c | c | c |} 
    	\hline
      	Datasets & Full Precision & XNOR\\
      	\hline
      	& Accuracy (\%) & Accuracy (\%)\\
      	\hline
      	CIFAR-10 & 93.29 & 80.24 \\
      	\hline
      	CIFAR-20 & 85.70 & 69.45 \\
      	\hline
      	CIFAR-30 & 80.87 & 60.90 \\
      	\hline
      	CIFAR-40 & 76.67 & 57.00 \\
      	\hline
      	CIFAR-50 & 75.18 & 54.36 \\
      	\hline
      	CIFAR-60 & 72.28 & 49.77 \\
      	\hline
      	CIFAR-70 & 72.70 & 51.20 \\
      	\hline
      	CIFAR-80 & 71.81 & 49.51 \\
      	\hline
      	CIFAR-90 & 71.61 & 50.22 \\
      	\hline
      	CIFAR-100 & 71.49 & 49.56 \\
      	\hline
    \end{tabular}}
    \label{tab:resnet18-cifar}
%   \vspace*{-3mm}
\end{table}

% \section{Summary}
\section{Conclusion}
\label{sec:summary}
This paper presents the first study of XNOR-Net for end-to-end classification of raw audio. We emphasize state-of-the-art MCUs as practically important target architectures for realistic Edge-AI applications.

For small problem sizes, measured in the number of classes, our comprehensive experimental analysis shows that XNOR-Net can produce models with reasonable performance that are sufficiently small to fit the target architectures. For our test scenarios with small problem sizes, XNOR networks require significantly less computation than comparably sized models generated by pruning-and-quantization alone. For relatively simple problems, the performance of XNOR-Nets on MCUs can be further increased by using spectrograms as input to the net. This enables us to capitalize on special DSP hardware so that the full-precision computations required to handle raw audio can be confined outside of the net in the spectrogram generation.\ignore{All of this makes XNOR well-suited for MCU targets and problems with few classes.} 

The picture changes as the problem complexity grows and larger numbers of classes need to be distinguished. Our analysis shows that XNOR-Nets face large drops in classification accuracy for datasets with many classes. This performance degradation is much more rapid than that of their pruning-and-quantization counterparts. This performance degradation can be handled by growing the network size. Although growing the network size may still allow the model to be run on CPUs, it takes away the ability to fit the model in  MCUs because XNOR-Net can only reduce the memory requirement by a maximum of 32-fold comparing to its full-precision version. Hence, for larger models, it is necessary to use other model compression techniques to find a model that is at most 32 times bigger than the target size before XNOR is applied.

In contrast, comparably sized models generated by pruning-and-quantization show significantly better performance while still satisfying the constraints of the target architectures. The advantages of spectrogram-based network input can no longer be exploited with complex data characteristics as feature-learning from raw audio or multi-channel input are required to reach state-of-the-art performance. This renders pruning-and-quantization the preferred approach from a certain problem complexity unless computation speed rather than model size dominates the decision.

Furthermore, to the best of our knowledge, there is no off-the-shelf computation kernel for XNOR-Nets yet. This means that it is difficult to realize the theoretical advantage of XNOR-Net on existing add-multiplication-based hardware and that custom hardware is required to achieve the full benefit of faster computation~\cite{jacob2018QuantTraining}. However, given the popularity of XNOR-Nets, we hope that such support is not too far away. 

Our study provides an experimental analysis of the current state of XNOR-Net and alternative compression methods as a basis to evaluate possible paths towards competitive deep learning architectures in edge-AI applications. The creation of specialized computation kernels for XNOR-Net and theoretical studies of XNOR-Net performance degradation for larger class sizes and how to avoid it are still open research problems and important areas for future research.

\bibliographystyle{unsrt}
\bibliography{ReferenceMaster}

\begin{thebibliography}{10}

\bibitem{der2019PredMaintenance}
Matthias~Auf der Mauer, Tristan Behrens, Mahdi Derakhshanmanesh, Christopher
  Hansen, and Stefan Muderack.
\newblock Applying sound-based analysis at porsche production: Towards
  predictive maintenance of production machines using deep learning and
  internet-of-things technology.
\newblock In {\em Digitalization Cases}, pages 79--97. Springer, 2019.

\bibitem{jia2018PredMaintenance}
Feng Jia, Yaguo Lei, Liang Guo, Jing Lin, and Saibo Xing.
\newblock A neural network constructed by deep learning technique and its
  application to intelligent fault diagnosis of machines.
\newblock {\em Neurocomputing}, 272:619--628, 2018.

\bibitem{yun2020PredMaintenance}
Huitaek Yun, Hanjun Kim, Eunseob Kim, and Martin~BG Jun.
\newblock Development of internal sound sensor using stethoscope and its
  applications for machine monitoring.
\newblock {\em Procedia Manufacturing}, 48:1072--1078, 2020.

\bibitem{sharan2016OverviewSoundRecognition}
Roneel~V Sharan and Tom~J Moir.
\newblock An overview of applications and advancements in automatic sound
  recognition.
\newblock {\em Neurocomputing}, 200:22--34, 2016.

\bibitem{xu2020CNNAecAnimalSpeech}
Weitao Xu, Xiang Zhang, Lina Yao, Wanli Xue, and Bo~Wei.
\newblock A multi-view cnn-based acoustic classification system for automatic
  animal species identification.
\newblock {\em Ad Hoc Networks}, 102:102115, 2020.

\bibitem{stowell2019SoundWildEcoMonitoring}
Dan Stowell, Tereza Petruskov{\'a}, Martin {\v{S}}{\'a}lek, and Pavel Linhart.
\newblock Automatic acoustic identification of individuals in multiple species:
  improving identification across recording conditions.
\newblock {\em Journal of the Royal Society Interface}, 16(153):20180940, 2019.

\bibitem{yan2019SoundWildEcoMonitoring}
Xiao Yan, Hemin Zhang, Desheng Li, Daifu Wu, Shiqiang Zhou, Mengmeng Sun,
  Haiping Hu, Xiaoqiang Liu, Shijie Mou, Shengshan He, et~al.
\newblock Acoustic recordings provide detailed information regarding the
  behavior of cryptic wildlife to support conservation translocations.
\newblock {\em Scientific reports}, 9(1):1--11, 2019.

\bibitem{gong2021AudioSpectroTransformer}
Yuan Gong, Yu-An Chung, and James Glass.
\newblock {AST: Audio Spectrogram Transformer}.
\newblock {\em arXiv preprint arXiv:2104.01778}, 2021.

\bibitem{nanni2021EnsenbleCNNAudio}
Loris Nanni, Gianluca Maguolo, Sheryl Brahnam, and Michelangelo Paci.
\newblock An ensemble of convolutional neural networks for audio
  classification.
\newblock {\em Applied Sciences}, 11(13):5796, 2021.

\bibitem{mohaimen2021ACDNet}
Md~Mohaimenuzzaman, Christoph Bergmeir, Ian~Thomas West, and Bernd Meyer.
\newblock Environmental sound classification on the edge: Deep acoustic
  networks for extremely resource-constrained devices.
\newblock {\em arXiv e-prints}, pages arXiv--2103, 2021.

\bibitem{kim2020MultiChannel}
Jaehun Kim.
\newblock Urban sound tagging using multi-channel audio feature with
  convolutional neural networks.
\newblock {\em Proceedings of the Detection and Classification of Acoustic
  Scenes and Events, 2020}, page
  \url{http://dcase.community/documents/challenge2020/technical_reports/DCASE2020_JHKim_21_t5.pdf},
  2020.

\bibitem{kumar2020SeqSelfTeaching}
Anurag Kumar and Vamsi Ithapu.
\newblock A sequential self teaching approach for improving generalization in
  sound event recognition.
\newblock In {\em Proceedings of the International Conference on Machine
  Learning, {ICML} 2020}, pages 5447--5457. PMLR, 2020.

\bibitem{zhang2019OnESC}
Zhichao Zhang, Shugong Xu, Shunqing Zhang, Tianhao Qiao, and Shan Cao.
\newblock Learning attentive representations for environmental sound
  classification.
\newblock {\em IEEE Access}, 7:130327--130339, 2019.

\bibitem{su2019Urbansound8k}
Yu~Su, Ke~Zhang, Jingyu Wang, and Kurosh Madani.
\newblock Environment sound classification using a two-stream {CNN} based on
  decision-level fusion.
\newblock {\em Sensors}, 19(7):1733, 2019.

\bibitem{tokozume2017Envnet2}
Yuji Tokozume, Yoshitaka Ushiku, and Tatsuya Harada.
\newblock Learning from between-class examples for deep sound recognition.
\newblock In {\em Proceedings of the 6th International Conference on Learning
  Representations, {ICLR} 2018}.
  \url{https://openreview.net/forum?id=B1Gi6LeRZ}, 2018.

\bibitem{sailor2017UnSupervisedFilterbank}
Hardik~B Sailor, Dharmesh~M Agrawal, and Hemant~A Patil.
\newblock Unsupervised filterbank learning using {Convolutional Restricted
  Boltzmann Machine} for environmental sound classification.
\newblock In {\em Proceedings of the 18th Annual Conference of the
  International Speech Communication Association, {Interspeech} 2017}, pages
  3107--3111, 2017.

\bibitem{han2015DeepCompression}
Song Han, Huizi Mao, and William~J. Dally.
\newblock Deep compression: Compressing deep neural network with pruning,
  trained quantization and huffman coding.
\newblock In {\em Proceedings of the 4th International Conference on Learning
  Representations, {ICLR} 2016}. \url{https://arxiv.org/abs/1510.00149}, 2016.

\bibitem{structuredpruning}
Xiaolong Ma, Geng Yuan, Sheng Lin, Zhengang Li, Hao Sun, and Yanzhi Wang.
\newblock {ResNet} can be pruned 60$\times$: Introducing network purification
  and unused path removal {(P-RM)} after weight pruning.
\newblock In {\em Proceedings of the IEEE/ACM International Symposium on
  Nanoscale Architectures, {NANOARCH} 2019}, pages 1--2. IEEE, 2019.

\bibitem{molchanov2016PrunningCNNNvdia}
Pavlo Molchanov, Stephen Tyree, Tero Karras, Timo Aila, and Jan Kautz.
\newblock Pruning convolutional neural networks for resource efficient
  inference.
\newblock In {\em Proceedings of the 5th International Conference on Learning
  Representations, {ICLR} 2017}.
  \url{https://openreview.net/forum?id=SJGCiw5gl}, 2017.

\bibitem{oyedotun2020SCompLasso}
Oyebade Oyedotun, Djamila Aouada, and Bjorn Ottersten.
\newblock Structured compression of deep neural networks with debiased elastic
  group {LASSO}.
\newblock In {\em Proceedings of the IEEE Winter Conference on Applications of
  Computer Vision, 2020}, pages 2277--2286, 2020.

\bibitem{hinton2015distilling}
Geoffrey Hinton, Oriol Vinyals, and Jeff Dean.
\newblock Distilling the knowledge in a neural network.
\newblock {\em stat}, 1050:9, 2015.

\bibitem{polino2018CompViaDistQuan}
Antonio Polino, Razvan Pascanu, and Dan Alistarh.
\newblock Model compression via distillation and quantization.
\newblock In {\em Proceedings of the 6th International Conference on Learning
  Representations, {ICLR} 2018}.
  \url{https://openreview.net/forum?id=S1XolQbRW}, 2018.

\bibitem{rastegari2016XNOR}
Mohammad Rastegari, Vicente Ordonez, Joseph Redmon, and Ali Farhadi.
\newblock {XNOR-Net}: Imagenet classification using binary convolutional neural
  networks.
\newblock In {\em Proceedings of the European conference on computer vision,
  {ECCV} 2016}, pages 525--542. Springer, 2016.

\bibitem{congXNORNet}
Cong Wang.
\newblock Pytorch {XNOR-Net}: {XNOR-Net}, with binary gemm and binary conv2d
  kernels, support both {CPU} and {GPU}.
\newblock \url{https://github.com/cooooorn/Pytorch-XNOR-Net}.
\newblock Accessed: Jun 15, 2021.

\bibitem{bulat2020ExpertBinNet}
Adrian Bulat, Brais Martinez, and Georgios Tzimiropoulos.
\newblock High-capacity expert binary networks.
\newblock In {\em Proceedings of the 9th International Conference on Learning
  Representations, {ICLR} 2021}.
  \url{https://openreview.net/forum?id=MxaY4FzOTa}, 2021.

\bibitem{cerutti2020AudioXnor}
Gianmarco Cerutti, Renzo Andri, Lukas Cavigelli, Elisabetta Farella, Michele
  Magno, and Luca Benini.
\newblock Sound event detection with binary neural networks on tightly
  power-constrained {IoT} devices.
\newblock In {\em Proceedings of the {ACM/IEEE} International Symposium on Low
  Power Electronics, {ISLPED} 2020}, pages 19--24, 2020.

\bibitem{rothmann2018WhatsWrongWithSpectroAsImage}
Daniel Rothmann.
\newblock What's wrong with spectrograms and {CNNs} for audio processing?
\newblock
  \url{https://towardsdatascience.com/whats-wrong-with-spectrograms-and-cnns-for-audio-processing-311377d7ccd},
  Mar 2018.

\bibitem{wyse2017SpectroByCNN}
L~Wyse.
\newblock Audio spectrogram representations for processing with convolutional
  neural networks.
\newblock In {\em Proceedings of the First International Conference on Deep
  Learning and Music, 2017}, pages 37--41, 2017.

\bibitem{chaudhary2021FFT}
Kartik Chaudhary.
\newblock Understanding audio data, fourier transform, {FFT}, spectrogram and
  speech recognition.
\newblock
  \url{https://towardsdatascience.com/understanding-audio-data-fourier-transform-fft-spectrogram-and-speech-recognition-a4072d228520},
  Jun 2021.

\bibitem{verma2018SpectroAsImage}
Prateek Verma and Julius~O Smith.
\newblock Neural style transfer for audio spectograms.
\newblock {\em arXiv preprint arXiv:1801.01589}, 2018.

\bibitem{irfan2021deepship}
Muhammad Irfan, Zheng Jiangbin, Shahid Ali, Muhammad Iqbal, Zafar Masood, and
  Umar Hamid.
\newblock Deepship: An underwater acoustic benchmark dataset and a separable
  convolution based autoencoder for classification.
\newblock {\em Expert Systems with Applications}, 183:115270, 2021.

\bibitem{ekpezu2021aecMfcc}
Akon~O Ekpezu, Isaac Wiafe, Ferdinand Katsriku, and Winfred Yaokumah.
\newblock Using deep learning for acoustic event classification: The case of
  natural disasters.
\newblock {\em The Journal of the Acoustical Society of America},
  149(4):2926--2935, 2021.

\bibitem{anwar2017StructuredPruningCNN}
Sajid Anwar, Kyuyeon Hwang, and Wonyong Sung.
\newblock Structured pruning of deep convolutional neural networks.
\newblock {\em ACM Journal on Emerging Technologies in Computing Systems
  (JETC)}, 13(3):32, 2017.

\bibitem{gordon2018MorphNet}
Ariel Gordon, Elad Eban, Ofir Nachum, Bo~Chen, Hao Wu, Tien-Ju Yang, and Edward
  Choi.
\newblock {MorphNet}: Fast \& simple resource-constrained structure learning of
  deep networks.
\newblock In {\em Proceedings of the IEEE Conference on Computer Vision and
  Pattern Recognition, {CVPR} 2018}, pages 1586--1595, 2018.

\bibitem{li2016PruningFiltersConvNet}
Hao Li, Asim Kadav, Igor Durdanovic, Hanan Samet, and Hans~Peter Graf.
\newblock Pruning filters for efficient convnets.
\newblock In {\em Proceedings of the 5th International Conference on Learning
  Representations, {ICLR} 2017}.
  \url{https://openreview.net/forum?id=rJqFGTslg}, 2017.

\bibitem{luo2018ThiNet}
Jian-Hao Luo, Hao Zhang, Hong-Yu Zhou, Chen-Wei Xie, Jianxin Wu, and Weiyao
  Lin.
\newblock {ThiNet}: pruning {CNN} filters for a thinner net.
\newblock {\em IEEE transactions on pattern analysis and machine intelligence},
  41(10):2525--2538, 2019.

\bibitem{singh2020SCompFilterCorel}
Pravendra Singh, Vinay~Kumar Verma, Piyush Rai, and Vinay Namboodiri.
\newblock Leveraging filter correlations for deep model compression.
\newblock In {\em Proceedings of the IEEE Winter Conference on Applications of
  Computer Vision, 2020}, pages 835--844, 2020.

\bibitem{luo2020autopruner}
Jian-Hao Luo and Jianxin Wu.
\newblock {AutoPruner}: An end-to-end trainable filter pruning method for
  efficient deep model inference.
\newblock {\em Pattern Recognition}, 107:107461, 2020.

\bibitem{piczak2015ESC}
Karol~J. Piczak.
\newblock {ESC}: {Dataset} for {Environmental Sound Classification}.
\newblock In {\em Proceedings of the 23rd {Annual ACM Conference} on
  {Multimedia}, 2015}, pages 1015--1018. {ACM Press}, 2015.

\bibitem{salamon2014Urbansound8k}
Justin Salamon, Christopher Jacoby, and Juan~Pablo Bello.
\newblock A dataset and taxonomy for urban sound research.
\newblock In {\em Proceedings of the 22nd ACM international conference on
  Multimedia, 2014}, pages 1041--1044. ACM, 2014.

\bibitem{huang2018Aclnet}
Jonathan~J Huang and Juan Jose~Alvarado Leanos.
\newblock {AclNet}: efficient end-to-end audio classification {CNN}.
\newblock {\em arXiv preprint arXiv:1811.06669}, 2018.

\bibitem{liu2018BiRealNet}
Zechun Liu, Baoyuan Wu, Wenhan Luo, Xin Yang, Wei Liu, and Kwang-Ting Cheng.
\newblock {Bi-Real Net}: Enhancing the performance of 1-bit {CNNs} with
  improved representational capability and advanced training algorithm.
\newblock In {\em Proceedings of the European conference on computer vision,
  {ECCV} 2018}, pages 722--737, 2018.

\bibitem{takahashi2016DNNAndDataAugmentationAcoustic}
Naoya Takahashi, Michael Gygli, Beat Pfister, and Luc Van~Gool.
\newblock Deep convolutional neural networks and data augmentation for acoustic
  event detection.
\newblock {\em arXiv preprint arXiv:1604.07160}, 2016.

\bibitem{meyer2017CNNForAED}
Matthias Meyer, Lukas Cavigelli, and Lothar Thiele.
\newblock Efficient convolutional neural network for audio event detection.
\newblock {\em arXiv preprint arXiv:1709.09888}, 2017.

\bibitem{Byerly2021NoRouting}
Adam Byerly, Tatiana Kalganova, and Ian Dear.
\newblock {No Routing Needed Between Capsules}.
\newblock {\em arXiv preprint arXiv:2001.09136v6}, 2021.

\bibitem{dosovitskiy2020image}
Alexey Dosovitskiy, Lucas Beyer, Alexander Kolesnikov, Dirk Weissenborn,
  Xiaohua Zhai, Thomas Unterthiner, Mostafa Dehghani, Matthias Minderer, Georg
  Heigold, Sylvain Gelly, et~al.
\newblock An image is worth 16x16 words: Transformers for image recognition at
  scale.
\newblock {\em arXiv preprint arXiv:2010.11929}, 2020.

\bibitem{tan2019efficientnet}
Mingxing Tan and Quoc Le.
\newblock {EfficientNet}: Rethinking model scaling for convolutional neural
  networks.
\newblock In {\em Proceedings of the International Conference on Machine
  Learning, {ICML} 2019}, pages 6105--6114. PMLR, 2019.

\bibitem{zhai2021scaling}
Xiaohua Zhai, Alexander Kolesnikov, Neil Houlsby, and Lucas Beyer.
\newblock Scaling vision transformers.
\newblock {\em arXiv preprint arXiv:2106.04560}, 2021.

\bibitem{he2016ResNet}
Kaiming He, Xiangyu Zhang, Shaoqing Ren, and Jian Sun.
\newblock Deep residual learning for image recognition.
\newblock In {\em Proceedings of the IEEE conference on computer vision and
  pattern recognition, {CVPR} 2016}, pages 770--778, 2016.

\bibitem{jacob2018QuantTraining}
Benoit Jacob, Skirmantas Kligys, Bo~Chen, Menglong Zhu, Matthew Tang, Andrew
  Howard, Hartwig Adam, and Dmitry Kalenichenko.
\newblock Quantization and training of neural networks for efficient
  integer-arithmetic-only inference.
\newblock In {\em Proceedings of the IEEE conference on computer vision and
  pattern recognition, {CVPR} 2018}, pages 2704--2713, 2018.

\end{thebibliography}

% \clearpage
\vskip -0.2\baselineskip plus -1fil
\begin{IEEEbiography}[{\includegraphics[width=1in,height=1.25in,clip,keepaspectratio]{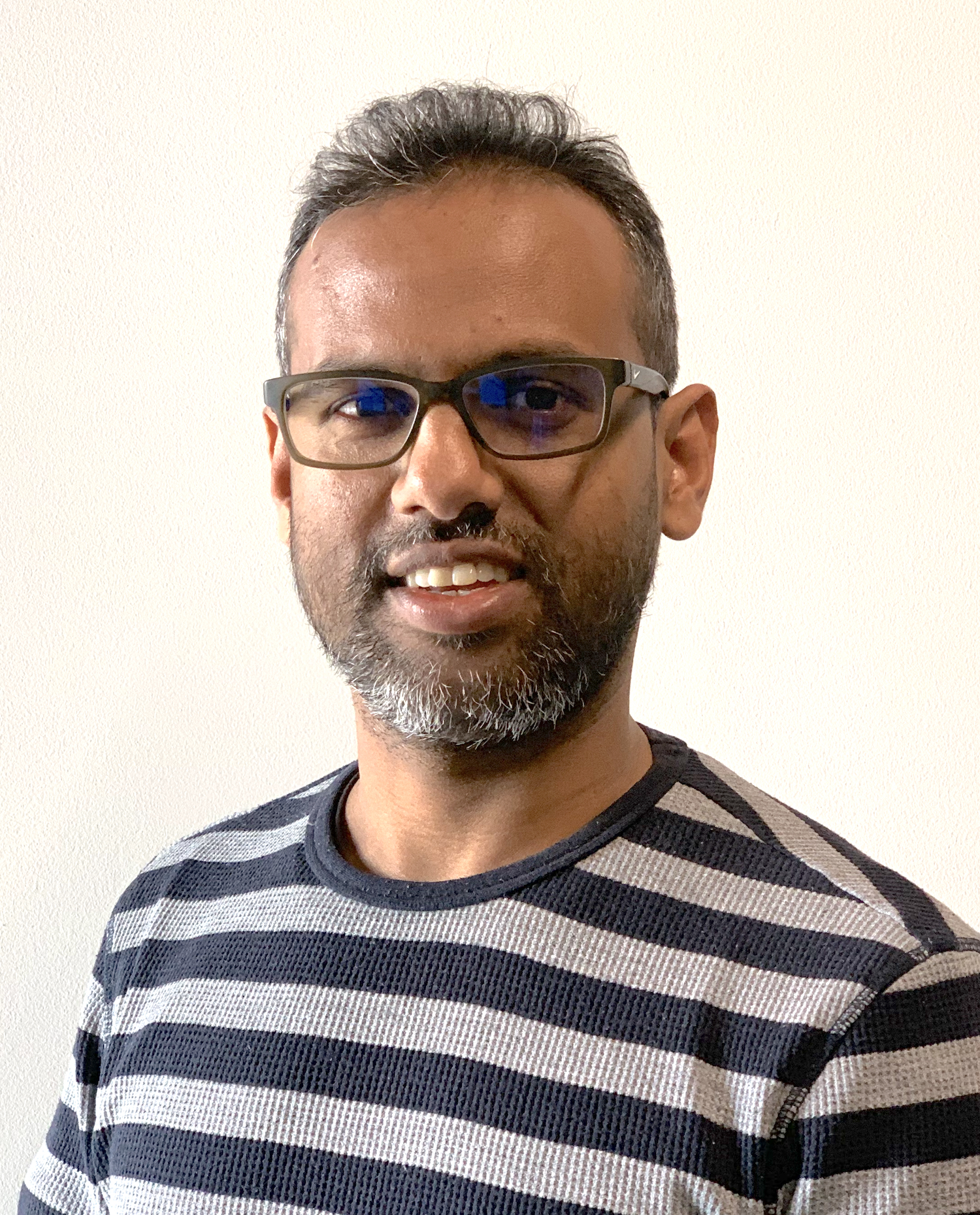}}]{\textbf{Md Mohaimenuzzaman}} received the B.Sc. and M.Sc. degrees in computer science and engineering, in 2007 and 2013, respectively. He is currently pursuing the Ph.D. degree in data science with the Department of Data Science and AI, Faculty of Information Technology, Monash University, Australia.

Before starting Ph.D., he developed software applications for international clients for about a decade. He is currently working as a Sessional Teaching Associate with the Faculty of Information Technology, where he teaches data science and software engineering related courses. He received the ‘‘2020 Faculty Teaching Excellence’’ award for teaching ‘‘Introduction to Data Science’’ to graduate students.
\end{IEEEbiography}

\vskip -0.2\baselineskip plus -1fil

\begin{IEEEbiography}[{\includegraphics[width=1in,height=1.25in,clip,keepaspectratio]{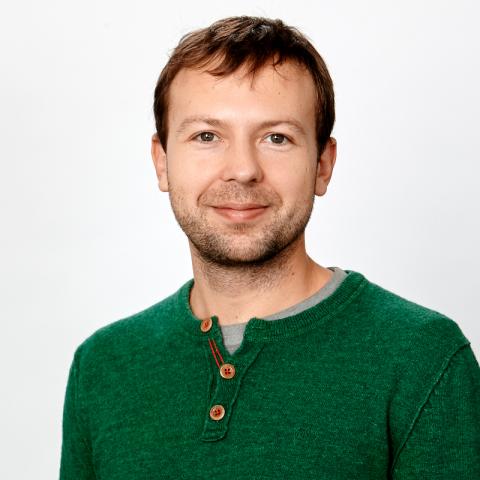}}]{\textbf{Christoph Bergmeir}} received the Ph.D. degree in computer science from the University of Granada, Spain, and the M.Sc. degree in computer science from the University of Ulm, Germany. 

He is currently a Senior Lecturer in data science and artiﬁcial intelligence with Monash University. He is a 2019 ARC DECRA Fellow with the Department of Data Science and AI, Monash University, where he develops ‘‘efﬁcient and effective analytics for real-world time series forecasting.’’ He works as a Data Scientist in a variety of projects with external partners in diverse sectors, e.g., in healthcare or infrastructure maintenance. He has published on time series prediction using machine learning methods, recurrent neural networks and long short-term memory neural networks (LSTM), time series predictor evaluation, and on medical applications and software packages in the R programming language, in journals, such as IEEE TRANSACTIONS ON NEURAL NETWORKS and LEARNING SYSTEMS, Journal of Statistical Software, Computational Statistics \& Data Analysis, and Information Sciences.
\end{IEEEbiography}

\vskip -0.2\baselineskip plus -1fil

\begin{IEEEbiography}[{\includegraphics[width=1in,height=1.25in,clip,keepaspectratio]{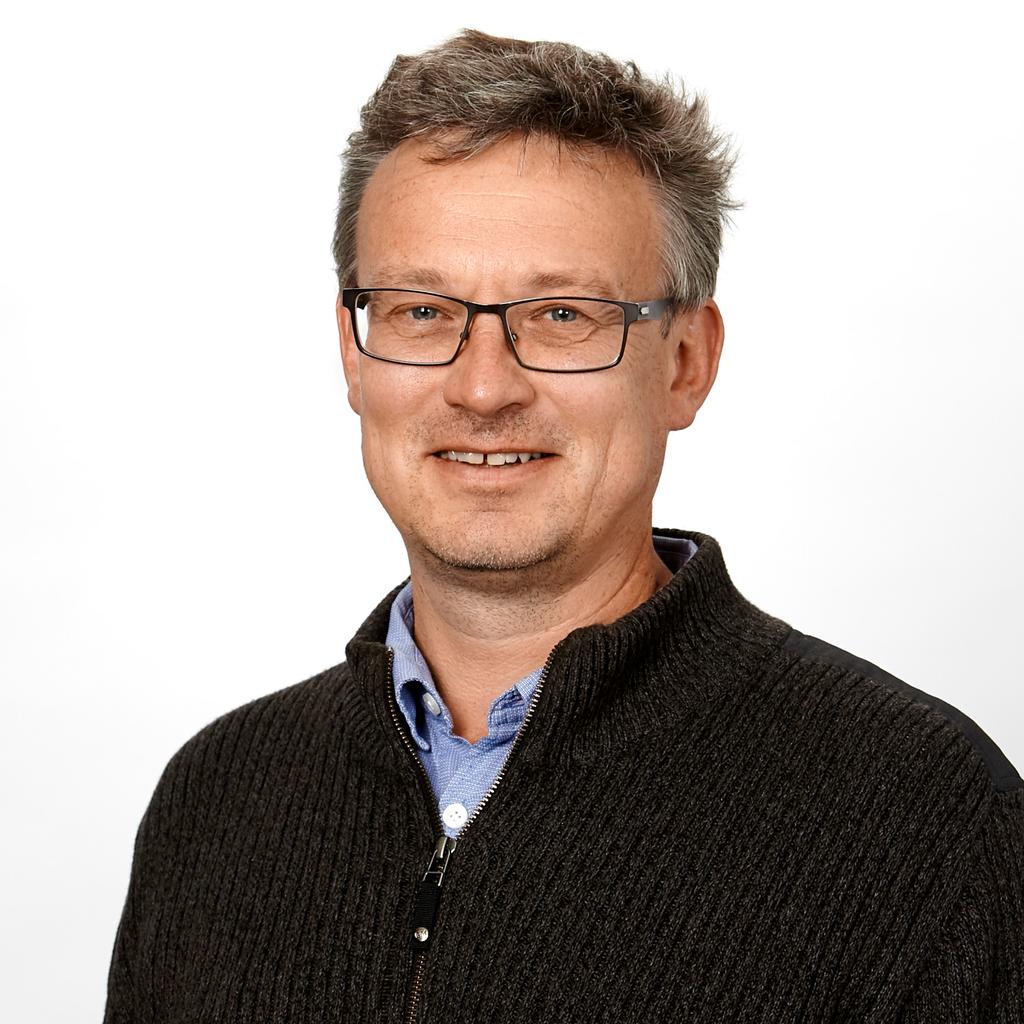}}]{\textbf{Bernd Meyer}} received the Ph.D. degree in computer science from the University of Hagen, Germany, in 1994. He is currently a Professor with the Department of Data Science and AI, Faculty of Information Technology, Monash University, Australia.

He works on data-intensive computational ecology, develops mathematical and computational models for the interactions of organisms with their environment, mostly focusing on the collective behaviour of social insects, such as bees and ants. How these self-organised ‘‘super-organisms’’ coordinate their actions remains a fascinating enigma. He also works on AI-based methods for monitoring animal activity as the basis for ecosystem monitoring and for automating experiments.
\end{IEEEbiography}

\EOD

\end{document}